\documentclass[journal]{IEEEtran}[12pt]
\ifCLASSINFOpdf
  \usepackage[pdftex]{graphicx}
  \graphicspath{{../pdf/}{../jpeg/}}
  \DeclareGraphicsExtensions{.pdf,.jpeg,.png}
\else
  \usepackage[dvips]{graphicx}
  \graphicspath{{../eps/}}
  \DeclareGraphicsExtensions{.pdf}
\fi
\usepackage{epstopdf}
\usepackage{stfloats}
\usepackage[cmex10]{amsmath}
\usepackage{algorithmic}
\usepackage{array}
\usepackage{mdwmath}
\usepackage{mdwtab}
\usepackage{graphicx}
\usepackage{subfigure}
\usepackage{setspace}
\usepackage{amssymb,amsmath,amsthm,amsfonts}
\usepackage{color}
\usepackage{hyperref} 
\usepackage{threeparttable}
\newtheorem{remark}{Remark} 
\newtheorem{lemma}{Lemma}

\begin{document}

\title{On Secure Mixed RF-FSO Systems With TAS and Imperfect CSI\thanks{Manuscript received.}}

\author{Hongjiang~Lei, ~\IEEEmembership{Senior Member,~IEEE,}
        Haolun~Luo,
        Ki-Hong~Park, ~\IEEEmembership{Member,~IEEE,}\\
        Imran~Shafique~Ansari, ~\IEEEmembership{Member,~IEEE,}
        Weijia~Lei,
        Gaofeng~Pan, ~\IEEEmembership{Senior Member,~IEEE,}
        and~Mohamed-Slim~Alouini, ~\IEEEmembership{Fellow,~IEEE}
\thanks{This work was supported by the National Natural Science Foundation of China under Grant 61971080, Chongqing Natural Science Foundation Project under Grant cstc2019jcyj-msxm1354, and the Open Fund of the Shaanxi Key Laboratory of Information Communication Network and Security under Grant ICNS201807.
}
\thanks{H. Lei, H. Luo, and W. Lei are with the School of Communication and Information Engineering, Chongqing University of Posts and Telecommunications, Chongqing 400065, China, also with Chongqing Key Lab of Mobile Communications Technology, Chongqing 400065, China. H. Lei is also with Shaanxi Key Laboratory of Information Communication Network and Security, Xi'an University of Posts \& Telecommunications, Xi'an, Shaanxi 710121, China (e-mail: leihj@cqupt.edu.cn).}
\thanks{I. S. Ansari is with James Watt School of Engineering, University of Glasgow, Glasgow G12 8QQ, United Kingdom.}
\thanks{G. Pan is with the School of Information and Electronics Engineering, Beijing Institute of Technology, Beijing 100081, China.}
\thanks{K.-H. Park and M.-S. Alouini are with CEMSE Division, King Abdullah University of Science and Technology (KAUST), Thuwal 23955-6900, Saudi Arabia.}
}

\maketitle

\begin{abstract}
In this work, we analyze the secrecy outage performance of a dual-hop relay system composed of multiple-input-multiple-output radio-frequency (RF) links and a free-space optical (FSO) link while a multiple-antenna eavesdropper wiretaps the confidential information by decoding the received signals from the source node. {The channel state information (CSI) of the RF and FSO links is considered to be outdated and imprecise, respectively.} We propose four transmit antenna selection (TAS) schemes to enhance the secrecy performance of the considered systems. The secrecy outage performance with different TAS schemes is analyzed and the effects of misalignment and detection technology on the secrecy outage performance of mixed systems are studied. We derive the closed-form expressions for probability density function (PDF) and cumulative distribution function (CDF) over M\'alaga channel with imperfect CSI. Then the closed-form expressions for the CDF and PDF of the equivalent signal-to-noise ratio (SNR) at the legitimate receiver over Nakagami-$m$ and M\'alaga channels are derived.
Furthermore, the bound of the effective secrecy throughput (EST) with different TAS schemes are derived. Besides, the asymptotic results for EST are investigated by exploiting the unfolding of Meijer's $G$-function when the electrical SNR of FSO link approaches infinity. Finally, Monte-Carlo simulation results are presented to testify the correctness of the proposed analysis. The results illustrate that outdated CSI shows a strong effect on the secrecy performance of the mixed RF-FSO systems. In addition, increasing the number of antennas at the source cannot significantly enhance the secrecy performance of the considered systems.
\end{abstract}



\section{Introduction}
\label{sec:introduction}
\subsection{Background and Related Literature}

The increase in multimedia services not only leads to increase in the requirements for the transmission rate of communications but also the emergence of communication security \cite{LiY2018TM}.
Free space optical (FSO) communications are regarded as a hopeful solution operating in the unlicensed spectrum \cite{Khalighi2014Survey, ZhangJ2015JSAC}.
It can offer high data rates and be utilized for various applications such as fiber backup, enterprise/local area network connectivity, back-haul for wireless cellular networks, metropolitan area network extensions, redundant link, and disaster recovery.
Relaying is a state-of-the-art technology that has been widely utilized in many wireless communication systems \cite{ZhangZ2017MNA}-\cite{ChengF2019TCOM}. By utilizing relaying technology, the mixed radio frequency (RF)-FSO systems combine both the advantages of the RF and FSO communication technologies.
The common performance of mixed RF-FSO systems was analyzed and the closed-form expressions for outage probability (OP), average symbol error rate (ASER), and ergodic capacity (EC) were derived vastly in the open literature (See Table I in \cite{Nasab2016TWC}).
In addition, mixed RF-FSO systems with multiple antennas or multiple users have been analyzed in many works \cite{ZhangJ2015JLT} - \cite{Varshney2018TCCN}. For example, the common performance of mixed RF-FSO systems with fixed-gain and variable-gain relaying schemes was investigated in \cite{ZhangJ2015JLT} and \cite{YangL2015JSAC}, respectively. Varshney {\it et al.} analyzed cognitive multiple-input-multiple-output (MIMO) RF-FSO systems with  perfect and outdated channel state information (CSI) in \cite{Varshney2017CL} and \cite{Varshney2018TCCN}, respectively.

As introduced in \cite{Hyadi2016Access, XiongJ2015TIFS}, it is impossible to obtain the perfect CSI for wireless links in practical systems {because of channels estimation errors, feedback errors, and delay.}
The mixed RF-FSO systems with outdated CSI of RF links was considered in many works, such as \cite{Varshney2017JLT} - \cite{Balti2018TCOM}.
The performance of a mixed RF-FSO decode-and-forward (DF) cooperative system was analyzed in \cite{Varshney2017JLT} where MIMO with zero-forcing based linear receiver was employed over RF links.
Salhab investigated the performance of a multiuser mixed RF-FSO relay network with generalized order user scheduling and obtained the closed-form expressions for OP, average bit error rate (ABER), and EC in \cite{Salhab2015AJSE}. Then a power allocation scheme was proposed to optimize the performance of the multiuser mixed RF-FSO system in \cite{Salhab2016JLT}.
The performance of mixed dual-hop RF-FSO system is analysed while variable gain relaying scheme was utilized at the relay node and the closed-form expressions for OP and ABER were derived in \cite{Djordjevic2015JSAC}.
Petkovic {\it et al.} analyzed the common performance of a dual-hop mixed RF-FSO system with partial amplify-and-forward (AF) relay selection scheme and derived the closed-form expressions for OP and ABER in \cite{Petkovic2015JLT}. A similar system was analyzed in \cite{Balti2018TCOM} wherein both AF and DF schemes were considered and the FSO link was assumed to follow M\' alaga distribution.
It must be noted that in \cite{Varshney2017JLT} - \cite{Balti2018TCOM}, the outdated CSI of RF links was considered and the CSI of FSO links was assumed to be precise. Furthermore, only the common performance metrics were investigated.

\begin{figure}[!t]
\centering{\includegraphics[width = 3.05 in]{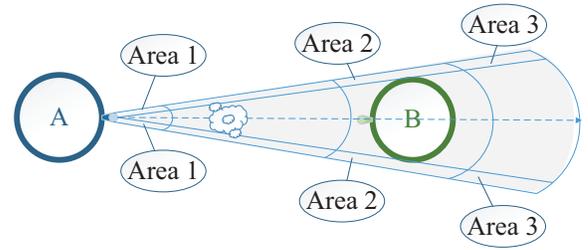}}
\caption{Wiretapping of a FSO link in the divergence region of the laser beam \cite{Lopez2015PJ, Monteiro2018TWC}.}
\label{FSOWyner}
\end{figure}

Recently, the physical layer security of the FSO or the mixed RF-FSO systems was investigated in many works \cite{Lopez2015PJ}-\cite{Lei2018TCOM}.
Due to the line-of-sight nature of FSO channels, such systems are considered to be highly secure.
However, in \cite{Lopez2015PJ}, Lopez-Martinez \emph{et al.} investigated the secrecy performance of FSO Wyner's model and derived the closed-form expressions for the probability of strictly positive secrecy capacity when the eavesdropper is near the transmitter or receiver, respectively.
The secrecy performance of coherent MIMO FSO systems with multi-aperture eavesdropper was investigated in \cite{Monteiro2018TWC} and the closed-form expressions for the effective secrecy throughput under transmit laser selection schemes were derived.
{In these two works, it is assumed that the eavesdropper is able to physically locate itself close to either the transmitter or the legitimate receiver, the areas 1 and 2 as shown in Fig. \ref{FSOWyner}.
Since the laser beam area at the receiver location is larger than the detector area at the receiver, some light propagates to the area behind the receiver. If the eavesdropper is located behind the receiver (the area 3 shown in Fig. \ref{FSOWyner}), it can still obtain the information.
}
The information security risk of a FSO system was investigated in \cite{ZouD2016PJ} wherein the eavesdropper could intercept the laser beam and hear the data signals through a non-line-of-sight scattering channel and the results showed the security concern is negligible when the visibility is clear enough.
Further, since it was stated in \cite{Lopez2015PJ} that they failed to imagine the practical design of eavesdroppers being able to operate in such scenarios, it was assumed that the eavesdropper can not intercept the laser beam transmitted by the relay \cite{AH2016TWC}-\cite{Lei2018TCOM}.
It must be noted that FSO link is secure, which is the main difference between FSO and RF links and also between mixed RF-FSO systems and traditional wireless communication systems.
El-Malek {\it et al.} analyzed the security reliability trade-off (SRT) of a multiuser mixed RF-FSO system in \cite{AH2016TWC} and derived the closed-form expressions for OP, ASER, EC, and intercept probability.
The effect on SRT of multiuser mixed RF-FSO systems from RF cochannel interference was analyzed in \cite{AH2017LT} and a power allocation scheme was proposed to enhance the secrecy performance of the considered system.
In previous work \cite{Lei2017PJ}, we analyzed the secrecy performance of the mixed RF-FSO uplink system wherein the closed-form expressions for secrecy outage probability (SOP) and average secrecy capacity (ASC) were derived under different relaying schemes and detection techniques.
The similar work was performed in \cite{YangL2018TVT}, wherein the RF and FSO links were assumed to experience with $\eta$-$\mu$ and {M\' alaga} distribution, respectively.
The secrecy outage performance of a mixed RF-FSO downlink systems was analyzed in \cite{Lei2018TCOM} where energy harvesting technology was utilized in RF links. We derived the expressions for the exact and asymptotic SOPs and discussed the effects of atmospheric turbulence, pointing error, detection technology, path loss, and energy harvesting on secrecy performance.

To the best of the authors' knowledge, no open literature addresses the performance of FSO systems with imprecise CSI except the recent studies \cite{FengJ2017OC} and \cite{Lei2018PJ}.
The common performance of FSO systems with imprecise CSI was analyzed and the closed-form expressions for the OP, ABER, and EC were derived in \cite{FengJ2017OC}. The atmospheric fading over FSO links was characterized by Gamma-Gamma turbulence model and only intensity modulation/direct detection (IM/DD) technology was considered.
Based on the results in \cite{FengJ2017OC}, the secrecy outage performance of a mixed single-input multiple-output (SIMO) RF-FSO system was analyzed in \cite{Lei2018PJ} and the imprecise CSI of both RF and FSO links was considered. But the RF and FSO links are modeled as Rayleigh and Gamma-Gamma channel, respectively.
The effect of estimation errors in both RF and FSO links on the secrecy performance of mixed systems was investigated in \cite{Lei2018PJ}.

\subsection{Motivation and Contributions}
M\' alaga distribution is the most generalized statistical model that describes the effect of irradiance fluctuation over FSO link.
It includes many other fading models for atmospheric optical communications, such as Log-Normal, Gamma-Gamma, Gamma-Rician, Shadowed-Rician, {\it etc.} as its special cases. Table I in \cite{JuradoA2011Malaga} lists the relationship between these existing distribution models and M\' alaga distribution.
{In this paper, the secrecy performance of the mixed RF-FSO systems with imperfect CSI\footnote{{In this work, it is assumed that the CSI of the RF links is outdated because of delay in feedback transmission and the CSI of the FSO link is imprecise due to channel estimation errors. The effect of delayed feedback coupled with estimation errors of CSI will be addressed in our future work.}} and four TAS schemes are proposed to improve the secrecy performance.
}
The main contributions of this article are:

\begin{itemize}
\item We analyze the statistical characteristics of M\' alaga fading channels with imprecise CSI under heterodyne detection (HD) and IM/DD techniques in the presence of pointing errors. The closed-form expressions for the probability density function (PDF) and cumulative distribution function (CDF) are obtained. To the best of author's knowledge based on the open literature, the results are new to the research community. One can easily utilize these to investigate the common performance metrics of FSO systems over M\' alaga fading channels with imprecise CSI. The PDF and CDF expressions given in \cite{FengJ2017OC} are a special case of this work.
\item {Four different TAS schemes, optimal transmit antenna selection (OTAS), TAS based on $S$-$R$ link (TASR), TAS based on $S$-$E$ link (TASE), and a new adaptive TAS scheme (ATAS)}, are proposed to enhance the secrecy performance of mixed RF-FSO systems. The secrecy outage performance of TASR and TASE schemes are analyzed and compared with the OTAS scheme. We derive the closed-form expressions for the exact and asymptotic EST/SOP for TASR and TASE schemes. Moreover, Monte-Carlo simulation results are demonstrated to validate the accuracy of our analytical results.
\item Differing from \cite{Varshney2017CL}, in which the TAS scheme was utilized to enhance the common performance while reducing the hardware complexity, in this work, we propose four different TAS schemes to improve the secrecy performance while the outdated CSI of RF links and imprecise CSI of the FSO link is considered.
\item Differing from \cite{Varshney2017JLT} - \cite{Balti2018TCOM}, we consider the outdated CSI for the RF links and the imprecise CSI for FSO links in this work wherein the respective secrecy performance is investigated. Technically speaking, it is much more challenging to analyze the secrecy performance relative to the common performance metrics, especially under outdated/imprecise CSI.
\item Differing from \cite{Lei2017PJ}, wherein all the nodes are equipped with a single antenna, the system assumption in this work is more practical since the source, relay, and eavesdropper are equipped with multiple antennas. Furthermore, both RF and FSO links are considered with outdated and imprecise CSI, respectively.
\item {Differing from \cite{Lei2018PJ}, the system in this work is generalized since the RF and FSO links are assumed to experience Nakagami-$m$ and  M\'alaga fading, respectively. Moveover, in \cite{Lei2018PJ}, the imprecise CSI of RF links is due to channel estimation errors, while in this work, the CSI of RF links is outdated because of the training/feedback delays.
    Last but not least, several TAS schemes are proposed to improve the secrecy performance of mixed RF-FSO systems.}
\end{itemize}

\subsection{Structure}
The remainder of this work is organized as follows. In Section \ref{sec:SystemModel}, we present the mixed RF-FSO system model and the statistical characteristics of each link are presented. The expressions of EST/SOP and its asymptotic results for different TAS schemes are obtained in Section \ref{sec:SOP}.
Monte-Carlo simulation and numerical results are presented in Section \ref{sec:RESULTS}. Finally, we conclude the work in Section \ref{sec:Conclusion}.

\section{System Model and Statistical Analysis}
\label{sec:SystemModel}
\begin{figure}[!t]
\centering{\includegraphics[width = 3.05 in]{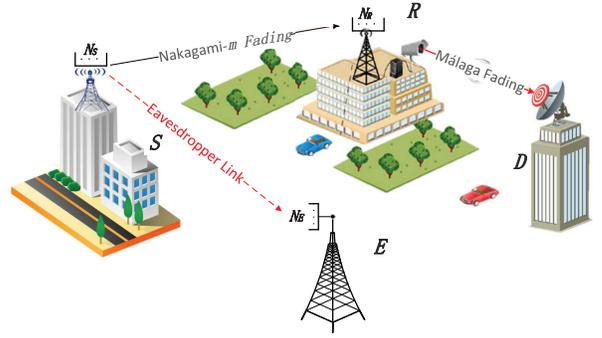}}
\caption{System model depicting the source $\left( S \right)$, relay $\left( R \right)$, destination $\left( D \right)$, and eavesdropper $\left( E \right)$ along with their respective antennas and links between them. The CSI of RF links are outdated and the CSI of the FSO link is imprecise.}
\label{systemmodel}
\end{figure}

Fig. \ref{systemmodel} shows a dual-hop mixed relay system composed of MIMO RF links and a FSO link.
A source node $\left( S \right)$ transmits the confidential information to a destination $\left( D \right)$ with the help of an  intermediate {DF} relay $\left( R \right)$ \footnote{The results in this work also fit with the bound of variable gain amplify-and-forward relaying scheme, as testified in vast existing literature, such as \cite{Lei2017PJ, Lei2018PJ, Zedini2016TWC, Zedini2015PJ}.}, while a multiple-antenna eavesdropper $\left( E \right)$ wiretaps the confidential information by decoding the received signals.
It is assumed that ${S}$ and ${E}$ are equipped with ${N_S}$ and ${N_E}$ RF antennas, respectively, $R$ is equipped with $N_R$ RF antennas and a transmit aperture, while $D$ is equipped with a single receive aperture. There is no direct link between $S$ and $D$ since there is no RF antenna on $D$.
TAS scheme is employed at ${S}$ since it makes full use of the advantages of MIMO system. Maximal ratio combining (MRC) scheme is utilized at both ${R}$ and ${E}$ to improve the received signal-to-noise ratio (SNR). The RF links experience quasi-static Nakagami-$m$ fading with parameter $m_i$ $\left( {i \in \left\{ {R,E} \right\}} \right)$ wherein the CSI is outdated and the FSO link follows quasi-static M\' alaga fading wherein the CSI is imprecise.
{Similar to \cite{AH2016TWC}-\cite{Lei2018TCOM}, it is assumed that $E$ is located beyond the divergence region of FSO link \footnote{{The scenario that $E$ is located in the divergence region of FSO link (inside of the shadow in Fig. \ref{FSOWyner}) will be considered as part of  the future works}.} (outside of the shadow in Fig. \ref{FSOWyner}), which means the FSO link is secure.
}

The {imprecise} FSO link channel gain ${\tilde h_{RD}}$ is expressed as \cite{FengJ2017OC}
\begin{equation}
{\tilde h_{RD}} = {\rho _{\text{FSO}}}{h_{RD}} + \sqrt {1 - \rho _{\text{FSO}}^2} \varepsilon,
\label{hRDhat}
\end{equation}
where {${{\tilde h}_{RD}}$ denotes the imperfect channel gain of $R$-$D$ link,
${h_{RD}}$ signifies the precise channel gain of $R$-$D$ link}, which is modeled as M\' alaga channel,
$\varepsilon $ is a Gaussian random variable that has zero-mean and unit variance and is independent of ${h_{RD}}$,
and {$ {\rho _{{\rm{FSO}}}}   \in \left( {0,1} \right]$ denotes the correlation coefficient.
A high value of ${\rho _{{\rm{FSO}}}}$ signifies low CSI estimation error, and ${\rho _{{\rm{FSO}}}} = 1$ denotes there is no estimation error. }

The instantaneous SNR of the FSO link can be expressed as  \cite{Ansari2016TWC}
\begin{equation}
{{{\tilde \gamma }_{RD}} = {{\bar \gamma }_{RD}}{\left( {{{\tilde h}_{RD}}} \right)^r},}
\label{gammaRD}
\end{equation}
where ${\bar \gamma _{RD}} $ is the average electrical SNR of FSO link, ${r}$ is the parameter that represents the type of detection technology being utilized, i.e. $r = 1$ accounts for HD and $r = 2$ represents IM/DD.

\begin{lemma}
\emph{The PDF and CDF of ${\tilde \gamma _{RD}}$ are obtained as
\begin{equation}
{f_{{{\hat \gamma }_{RD}}}}\left( \gamma  \right) = \left\{ {\begin{array}{*{20}{c}}
{{B_D}\sum\limits_{h = 1}^\beta  {\sum\limits_{k = 0}^\infty  {{b_h}{\varphi _1}{e^{ - {\psi _1}{\gamma ^{\frac{2}{r}}}}}} } {\gamma ^{\frac{{k + 1}}{r} - 1}},}&{\gamma  > 0}\\
{1 - {{\rm{Z}}_0}}&{\gamma  = 0}
\end{array}} \right.,
\label{pdfgammaRD}
\end{equation}
\begin{equation}
{F_{{{\hat \gamma }_{RD}}}}\left( \gamma  \right) = \left\{ {\begin{array}{*{20}{c}}
{{B_D}\sum\limits_{h = 1}^\beta  {\sum\limits_{k = 0}^\infty  {{H_1}} } G_{1,2}^{1,1}\left[ {{\psi _1}{\gamma ^{\frac{2}{r}}}\left| {_{\frac{{1 + k}}{2},0}^1} \right.} \right],}&{\gamma  > 0}\\
{1 - {{\rm{Z}}_0},}&{\gamma  = 0}
\end{array}} \right.,
\label{cdfgammaRD}
\end{equation}
respectively, where
$B _D = \frac{{{\xi ^2}{A_D}{2^{\alpha  - 4.5}}}}{{{\pi ^{1.5}}}}$,
${A_D} = \frac{{2{\alpha ^{0.5\alpha }}}}{{{g^{1 + 0.5\alpha }}\Gamma \left( \alpha  \right)}}{\left( {\frac{{g\beta }}{{g\beta  + {\Omega _1}}}} \right)^{\beta  + 0.5\alpha }}$,
${\varphi _1} = \frac{{{2^{0.5k + h}}{G_k}}}{{rk!{{\bar \gamma }_{RD}}^{\frac{{1 + k}}{r}}{{\left( {1 - \rho _{FSO}^{\rm{2}}} \right)}^{{\rm{0}}{\rm{.5}}\left( {k + 1} \right)}}}}$,
${b_h} = \frac{{\left( {\beta  - 1} \right)!{{\left( {g\beta  + {\Omega _1}} \right)}^{1 - 0.5h}}{\alpha ^{0.5h}}{\Omega _1}^{h - 1}}}{{{{\left( {\left( {h - 1} \right)!} \right)}^2}\left( {\beta  - h}! \right){\beta ^{0.5h}}{g^{h - 1}}}}{\left( {\frac{{\alpha \beta }}{{g\beta  + {\Omega _1}}}} \right)^{ - 0.5\left( {\alpha  + h} \right)}}$,
${\psi _1} = \frac{1}{{2{{\bar \gamma _{RD}}}^{\frac{2}{r}}\left( {1 - \rho _{\text{FSO}}^2} \right)}}$,
${H_1}{\text{ = }}\frac{{{b_h}{2^{k + h - 0.5}}{G_k}}}{{k!}}$,
${{\text{Z}}_0} = \frac{{{\xi ^2}{A_D}}}{{{\pi ^{1.5}}}}\sum\limits_{h = 1}^\beta  {\sum\limits_{k = 0}^\infty  {\frac{{{2^{k + \alpha  + h - 5}}{b_h}{G_k}}}{{k!}}\Gamma \left( {\frac{{k + 1}}{2}} \right)} }$,
$\Gamma \left(  \cdot  \right)$ is the Gamma function, as defined by \cite[(8.310)]{Gradshteyn2007},
${\alpha}$ and ${\beta}$ are fading parameters,
$g = 2{b_0}\left( {1 - {\rho _0}} \right)$ denotes the average power of the scattering component received by off-axis eddies,
$2{b_0}$ is the average power of the total scatter components, $\rho _0$  represents the amount of scattering power coupled to the line-of-sight (LOS) component,
$\Omega _1 = \Omega  + 2{b_0}{\rho _0} + 2\sqrt {2{b_0}{\rho _0}\ \Omega} \cos \left( {{\varphi _A} - {\varphi _B}} \right)$ represents the average power from the coherent contributions,
$\Omega $ is the average power of the LOS component, ${\varphi _A}$ and ${\varphi _B}$ are the deterministic phases of the LOS and the coupled-to-LOS scatter terms, respectively.
${G_k} = G_{6,3}^{1,6}\left[ {\frac{{{2^3}}}{{{\delta ^2}\left( {1 - {\rho _{\text{FSO}}}^2} \right)}}\left| {_{\frac{k}{2},\frac{{ - {\xi ^2}}}{2},\frac{{1 - {\xi ^2}}}{2}}^{\frac{{1 - {\xi ^2}}}{2},\frac{{2 - {\xi ^2}}}{2},\frac{{1 - \alpha }}{2},\frac{{2 - \alpha }}{2},\frac{{1 - h}}{2},\frac{{2 - h}}{2}}} \right.} \right]$, where $\delta  = \frac{{\alpha \beta }}{{\left( {g\beta  + \Omega _1} \right){I_l}{A_0}{\rho _{\text{FSO}}}}}$,
$I_l$ is the path loss that is a constant in a given weather condition,
$A_0$ and $\xi$ are constant terms that defines the pointing loss,
and $G_{p,q}^{m,n}\left[ \cdot \right]$ is the Meijer's $G$-function as defined by \cite[(9.301)]{Gradshteyn2007}.}
\end{lemma}

{\it Proof :} See Appendix A.

\begin{remark}
\emph{Table I in \cite{JuradoA2011Malaga} lists many existing distribution models that can be utilized for atmospheric optical communications and how these models can be generated from the M\' alaga distribution model. Thus one can easily obtain the PDF and CDF for all these models with imprecise CSI. For instance, when $\rho _0= 1$,
${\Omega _1} = 1$, $r = 1$, the CDF presented in (\ref{cdfgammaRD}) agrees with the individual result presented by (23) (for $\xi \to \infty $) and (24) of \cite{FengJ2017OC}, respectively. Based on the results in this work and making use of the method proposed in \cite{Zedini2016TWC} and \cite{Zedini2015PJ}, one can easily analyze the common performance metrics of the corresponding systems over M\' alaga distribution with imperfect CSI, such as OP, ABER/ASER, and EC.}
\end{remark}

\begin{remark}
\emph{{From (\ref{cdfgammaRD}), one can observe that there is a floor (equal to $1 - {{\text{Z}}_0}$) for the OP when ${\bar \gamma _{RD}} \to \infty $ because ${{\text{Z}}_0}$ is independent of ${\bar \gamma _{RD}} $. The basic reason is that ${{\tilde h}_{RD}}$ must be positive in practical models while $\varepsilon $  can be positive or negative.}}
\end{remark}

\section{Effective Secrecy Throughput Analysis}
\label{sec:SOP}

In this work, it is assumed that the main traffic is delay-intolerant transmission and
the EST is utilized as the performance metric. {For the delay-tolerant transmission scenarios, the source transmits at any constant rate upper bounded by the ergodic capacity. Since the codeword length is sufficiently large compared to the block time, the codeword could experience all possible realizations of the channel. As such, the ASC becomes an appropriate measure. Hence, the EST of the mixed system is given by
$\Psi  = {C_s}$, where ${C_s}$ is the ASC of the mixed system, which will be investigated in our future work based on the results of this work.}

The EST for the delay-intolerant traffic is defined as \cite{JiangX2016TCOM, ZhongC2015TCOM}
\footnote{Refs. \cite{YanS2015TWC} and \cite{GomezG2019Access} gave another definition of EST, which is $\Psi \left( {{R_{\rm{B}}},{R_{\rm{E}}}} \right) = {R_s}\left( {1 - {P_{{\rm{ro}}}}\left( {{R_{\rm{B}}}} \right)} \right)\left( {1 - {P_{{\rm{so}}}}\left( {{R_{\rm{E}}}} \right)} \right)$, where ${R_{\rm{B}}}$ dentes the codeword rate and ${R_{\rm{E}}} = {R_{\rm{B}}} - R_s$ denotes the redundancy rate, ${P_{{\rm{ro}}}}\left( {{R_{\rm{B}}}} \right) = \Pr \left( {{R_{\rm{B}}} > {C_{\rm{B}}}} \right)$ and ${P_{{\rm{so}}}}\left( {{R_{\rm{E}}}} \right) = \Pr \left( {{R_{\rm{E}}} < {C_{\rm{E}}}} \right)$ signify the reliability probability and intercept probability, respectively. The EST of the mixed system with this definition can be easily obtained with the aid of the results of this work. }
\begin{equation}
{\Psi  = {R_s}\left( {1 - {P_{out}}} \right),}
\label{EST}
\end{equation}
where $R_s$ signifies the constant transmission rate and ${P_{out}}$ denotes the SOP of the mixed systems, which is expressed as \cite{Bloch2008TIT}
\begin{equation}
\begin{aligned}
{P_{out}} & = \Pr \left\{ {{C_s}\left( {{\gamma _{eq}},{\gamma _{SE}}} \right) \leqslant {R_s}} \right\} \\
& = \int_0^\infty  {{F_{eq}}\left( {\Theta {\gamma _{SE}} + \Theta  - 1} \right)} {f_{{\gamma _{SE}}}}\left( {{\gamma _{SE}}} \right)d{\gamma _{SE}},
\label{Pout}
\end{aligned}
\end{equation}
where
${C_s}\left( {{\gamma _{eq}},{\gamma _{SE}}} \right) = {\left[ {{{\log }_2}\left( {1 + {\gamma _{eq}}} \right) - {{\log }_2}\left( {1 + {\gamma _{SE}}} \right)} \right]^ + }$ denotes the secrecy capacity of mixed RF-FSO systems,
${\left[ x \right]^ + } = \max \left\{ {x,0} \right\}$,
and $\Theta  = 2^{R_s}$.

\begin{remark}
{One can easily find from (\ref{Pout}) that a larger $R_s$ will lead to a larger ${P_{out}} $ and vice versa. This means there is an optimal $R_s$ to obtain the maximum EST for the corresponding scenario. Based on the results of this work, one can easily obtain the optimal $R_s$ and the maximum $\Psi$ by utilizing root-finding method \cite{YangN2012TCOM}.}
\end{remark}

In this work, we investigate the secrecy outage performance of the mixed RF-FSO system because of the following reasons:
1) In some cases, the source node can choose certain fixed rates within a limited range due to the constraint by the coding and modulation schemes even if the source node knows the CSI of all the links;
2) Since the CSI is known to be outdated/imprecise, the rate selected by the CSI may still cause outages due to the outdated/channel estimation errors.
3) Even if the global CSI is known, it is useful for the source node to evaluate the secrecy performance through SOP. In some cases, such as the delay-intolerant traffic scenario \cite{JiangX2016TCOM} and \cite{ZhongC2015TCOM}, the data rate above a certain constant rate is required at the cost of a certain amount of reliability (i.e. SOP) over transmission time when the channel is slowly varying.
In these scenarios, the transmitter transmits at a constant rate $R_s$ that is required in the system regardless of whatever rate in the channel is achieved. In other word, we can utilize the trade-off between reliability and rate.
We can achieve the higher constant rate than ASC at the expense of reliability if this reliability is acceptable in the system.
One can achieve perfect secrecy rate and, on average, this might be above the constant rate $R_s$. However, in the case of slow fading scenario, OP, which evaluates the systems probability of being robust to the outage, is a more important metric than the average secrecy capacity. Overall, it means that even though the perfect secrecy rate is guaranteed, the OP above a certain constant rate is also an important performance metric in some scenarios where no transmission case is also regarded as ``outage".

It must be noted that there is a Meijer's $G$-function presented in the CDF of FSO link, and hence obtaining a closed-form result for the integral including the shift in the Meijer's $G$-function is almost impossible and/or too complex.
Therefore in the subsequent section, the lower bound of the SOP is considered, which has been utilized in many works \cite{Lei2017PJ, Lei2018PJ}.
\begin{equation}
\begin{aligned}
P_{out}^{\text{L}} &= \Pr \left\{ {{\gamma _{eq}} \leqslant \Theta {\gamma _{SE}}} \right\}\\
& = \int_0^\infty  {{F_{eq}}\left( {\Theta {\gamma _{SE}} } \right)} {f_{{\gamma _{SE}}}}\left( {{\gamma _{SE}}} \right)d{\gamma _{SE}}.
\label{PoutL}
\end{aligned}
\end{equation}

\begin{remark}
Compared with (\ref{Pout}) and (\ref{PoutL}), one can find that the tightness between the exact SOP and lower bound for SOP depend on $\Theta  = {2^{{R_s}}}$ and ${\gamma _{SE}}$. We can easily observe that the smaller $R_s$ and/or larger ${\bar \gamma _{SE}}$, the tighter the bound.
\end{remark}
%

In the following of this section, $P_{out}^{\text{L}}$ for different TAS schemes are investigated carefully. Substituting these results into (\ref{EST}), the analytical expressions for the bound of the EST and the asymptotic EST can be easily obtained.

\subsection{SOP with the `Optimal' Transmit Antenna Selection}
In the scenario where the source node has the global CSI (a general assumption in the physical layer  security literature), the antenna can be selected appropriately in order to maximize the secrecy performance of the mixed RF-FSO system.
Since larger instantaneous secrecy capacity always signifies better secrecy outage performance, we named this scheme as OTAS\footnote{{
It should be noted that in order to obtain the real optimal secrecy performance, the transmit antenna must be selected based on the accurate CSI. In other words, the OTAS scheme, which is based on the available (imperfect) CSI, can not necessarily obtain the optimal secrecy capacity during transmission within the considered system due to the imperfect CSI. The result of OTAS is optimal from the user's point of view.}}.
Based on the results in \cite{Lei2017TVT}, the following criterion  can be utilized
\begin{equation}
{b_{{\rm{OTAS}}}} = \arg \mathop {\max }\limits_{1 \le i \le {N_S}} \left( {{{ C_{s,i}^{{\rm{OTAS}}}}}} \right),
\label{bOTAS}
\end{equation}
where ${{ C_{s,i}^{{\rm{OTAS}}}} = \log \left( {\frac{{1 + \min \left( {{ \gamma _{{S_i}R}},{{ \gamma }_{RD}}} \right)}}{{1 + {\hat \gamma _{{S_i}E}}}}} \right)}$ signifies the secrecy capacity of a mixed RF-FSO system when the $i$-th antenna at $S$ is selected.
Then we can express the SOP under OTAS scheme as
\begin{equation}
\begin{aligned}
P_{out}^{{\rm{OTAS}}} &= \Pr \left\{ { C_{s,{b_{{\rm{OTAS}}}}}^{{\rm{OTAS}}} \le {R_s}} \right\}\\
& = \Pr \left\{ {\max \left( { C_{s,i}^{{\rm{OTAS}}}} \right) \le {R_s}} \right\}\\
& = \Pr \left\{ { C_{s,1}^{{\rm{OTAS}}} \le {R_s}, \cdots,  C_{s,{N_S}}^{{\rm{OTAS}}} \le {R_s}} \right\}.
\label{sopOAS}
\end{aligned}
\end{equation}

{However, it is challenging to obtain the closed-form solutions for SOP under OTAS because ${\hat C_{s,i}^{{\rm{OTAS}}}}$ are not independent of each other. Another important reason is that it is very difficult to obtain the exact CSI of $R$-$D$ for $S$ due to delay in feedback transmission of the first hop.
In Section \ref{sec:RESULTS}, simulation results of SOP under OTAS are presented as a benchmark with the assumption that $S$ has the imperfect CSI of FSO link without delay.}

\begin{remark}
Since only partial links are considered in selection metrics, the secrecy performance under TASR and TASE schemes will not exceed that under OTAS.
\end{remark}

In the following, three different suboptimal TAS schemes are proposed to enhance the secrecy performance of the mixed RF-FSO systems.

\subsection{SOP with TAS based on $S$-$R$ link}
In proactive eavesdropping scenarios where the eavesdropper's CSI is unknown at $S$,
one can select an antenna based on the CSI of $S$-$R$ links.
The TAS criterion of this scheme (TASR) is expressed as
\begin{equation}
{b_1} = \arg \mathop {\max }\limits_{1 \le i \le {N_S}} \left( {\sum\limits_{j = 1}^{{N_R}} {{{\left| {{ h_{{S_i}{R_j}}}} \right|}^2}} } \right),
\label{bR}
\end{equation}
where $b_1$ signifies the selected antenna at $S$ with this scheme. Thus the actual SNR at $R$ during data transmission can be written as
\begin{equation}
{{\hat \gamma }_{SR, 1}} = \frac{{{P_S}}}{{{\sigma ^2}}}\sum\limits_{j = 1}^{{N_R}} {{{\left| {{{ \hat h}_{{S_{b_1}}{R_j}}}} \right|}^2}},
\label{gammaSRhat}
\end{equation}
where $P_S$ is the transmit power at $S$ and ${\sigma ^2}$ denotes the variance of additive white Gaussian noise (AWGN) {and ${\hat h_{{S_{{b_1}}}{R_j}}}$ is the time-delayed version of ${h_{{S_{{b_1}}}{R_j}}}$.}

The correlation relationship between the outdated and accurate channel gain can be expressed as \cite{Ferdinand2013CL}
\begin{equation}
{\hat h_{{S_i}{R_j}}} = {\rho _{SR}} {h_{{S_i}{R_j}}} + \sqrt {1 - {\rho _{SR} ^2}} \omega _1,
\label{hsrhat}
\end{equation}
where {${h_{{S_i}{R_j}}}$ and ${\hat h_{{S_i}{R_j}}}$ are the accurate and outdated channel gain} between the $i$-th antenna at $S$ and $j$-th antenna at $R$, respectively, $\omega _1$ represents a Nakagami-$m$ random variable with same variance as ${h_{{S_i}{R_j}}}$, and $0 < {\rho _{SR}} < 1$ is the correlation coefficient.

\begin{remark}
It should be noted that there are two main differences between (\ref{hRDhat}) and (\ref{hsrhat}), although they are expressed in similar expression. Firstly, ${\hat h_{RD}}$ must be positive real value while ${\hat h_{{S_i}{R_j}}}$ can be any complex value. Secondly, $\varepsilon $ signifies the channel estimation errors while $\omega _1$ represents the errors from being outdated.
For the case that $N_S = 1$, $\omega _1$ will not influence ${\hat h_{{S_i}{R_j}}}$ since there is no antenna selection at $S$.
\end{remark}


\begin{lemma}
\emph{The PDF and CDF of ${\hat \gamma _{SR,1}} $ can be expressed as
\begin{equation}
{f_{{\hat \gamma _{SR,1}}}}\left( \gamma  \right) = {\phi _R}\sum\limits_{S _R} {\sum\limits_{q = 0}^{B _R} {{\Lambda _R}{\gamma ^{q + {\tau _R} - 1}}{e^{ - {\upsilon _R}\gamma }}} },
\label{pdfgammasrhat}
\end{equation}
\begin{equation}
{F_{{\hat \gamma _{SR,1}}}}\left( \gamma  \right) = 1 - {\phi _R}\sum\limits_{S _R} {\sum\limits_{q = 0}^{B _R} {\sum\limits_{t = 0}^{q + {\tau _R} - 1} {{\varphi _R}{\gamma ^t}{e^{ - {\upsilon _R}\gamma }}} } },
\label{cdfgammasrhat}
\end{equation}
respectively, where ${\phi _R} = \frac{{{N_S}\lambda _R^{{\tau _R} + 1}\rho _{SR}^{1 - {\tau _R}}}}{{\left( {1 - \rho _{SR}^2} \right)\Gamma \left( {{\tau _R}} \right)}}$,
${\tau _R} = {N_R}{m_R}$,
${\lambda _R} = \frac{{{m_R}}}{{{{\bar \gamma }_{SR}}}}$,
${B _R} = \sum\limits_{p = 2}^{{\tau _R} + 1} {{n_p}\left( {p - 2} \right)} $,
${S_R} = \left\{ {\left( {{n_1}, \cdots ,{n_{{\tau _R} + 1}}} \right) \in \mathbb{N}\left| {\sum\limits_{p = 1}^{{\tau _R} + 1} {{n_p} = {N_S} - 1} } \right.} \right\}$,
$\mathbb{N}$ denotes a non-negative integer set,
${\upsilon _R} = {\frac{{{\lambda _R}}}{{1 - \rho _{SR}^2}} - \frac{{{{\left( {\lambda _R \beta _R} \right)}^2}}}{\alpha _R}} $,
${\Lambda _R} = \frac{{{A _R}{B _R}!{{\left( {{\lambda _R}\beta _R} \right)}^{2q + {\tau _R} - 1}}\left( {{B _R} + {\tau _R} - 1} \right)!}}{{q!{\alpha _R^{{B _R} + {\tau _R} + q}}\left( {{B _R} - q} \right)!\left( {{\tau _R} + q - 1} \right)!}}$,
${\varphi _R} = \frac{{{\Lambda _R}\left( {q + {\tau _R} - 1} \right)!{\upsilon _R^{t - q - {\tau _R}}}}}{{t!}}$,
${\beta _R} = \frac{{ {{\rho _{SR}}} }}{{1 - \rho _{SR}^2}} $,
${A_R} = \left( {\frac{{{N_S} - 1}}{{\prod\limits_{q = 1}^{{\tau _R} + 1} {{n_q}} }}} \right)\prod\limits_{p = 2}^{{\tau _R} + 1} {{{\left( { - \frac{{\lambda _R^{p - 2}}}{{\left( {p - 2} \right)!}}} \right)}^{{n_p}}}} $,
${\alpha _R} = {\lambda _R}\left( {\frac{{{\rho _{SR}^2}}}{{1 - \rho _{SR}^2}} + {C_R} + 1} \right) $,
${C _R} = \sum\limits_{p = 2}^{{\tau _R} + 1} {{n_p}} $,
and
${\bar \gamma _{SR}}$ is the average SNR of ${S-R}$ link.}
\end{lemma}

{\it {Proof :}} See Appendix B.

Since DF relaying scheme is utilized at $R$, the equivalent SNR at $D$ can be expressed as \cite{Lei2018TCOM}
\begin{equation}
\gamma _{eq} =  \min \left( {{\hat \gamma _{SR,1}},{\tilde \gamma _{RD}}} \right).
\label{gammaeq}
\end{equation}
The CDF of $\gamma _{eq}$ in this case is obtained as
\begin{equation}
\begin{aligned}
{F_{\gamma _{eq, 1}}}\left( \gamma  \right) &= {F_{{{\hat \gamma }_{SR}}}}\left( \gamma  \right) + {F_{{{\hat \gamma }_{RD}}}}\left( \gamma  \right) - {F_{{{\hat \gamma }_{SR}}}}\left( \gamma  \right){F_{{{\hat \gamma }_{RD}}}}\left( \gamma  \right)\\
& = 1 + {\phi _R}\sum\limits_{S _R} {\sum\limits_{q = 0}^{B _R} {\sum\limits_{t = 0}^{q + {\tau _R} - 1} {{\varphi _R}{\gamma^t}{e^{ - \upsilon _R \gamma}}} } } \\
& \times  \hfill \left( {B _D\sum\limits_{h = 1}^\beta  {\sum\limits_{k = 0}^\infty  {{H_1}} } G_{1,2}^{1,1}\left[ {{\psi _1}{\gamma ^{\frac{2}{r}}}\left| {_{\frac{{1 + k}}{2},0}^1} \right.} \right] - {{\text{Z}}_0}} \right).
\label{cdfgammaeq1}
\end{aligned}
\end{equation}

TAS is based on the CSI of $S$-$R$ link, which means a random transmit antenna would be selected for $E$ \cite{Lei2017TVT}. Then the SNR at $E$ in this case can be expressed as
\begin{equation}
{\gamma _{SE, 1}} = \frac{{{P_S}}}{{{\sigma ^2}}}\sum\limits_{j = 1}^{{N_E}} {{{\left| {{h_{{S_{b_1}}{E_j}}}} \right|}^2}}.
\label{gammase1}
\end{equation}

The PDF and CDF of ${\gamma _{SE, 1}} $ are given by \cite{Lei2017TVT}
\begin{equation}
{f_{{\gamma _{SE, 1}}}}\left( x \right) = \frac{{{\lambda _E^{{\tau _E}}}}}{{\Gamma \left( {{\tau _E}} \right)}}{e^{ - {\lambda _E}x}}{x^{{\tau _E} - 1}},
\label{pdfgammase1}
\end{equation}
\begin{equation}
{F_{{\gamma _{SE, 1}}}}\left( x \right) = 1 - \sum\limits_{i = 0}^{{\tau _E} - 1} {\frac{{{e^{ - {\lambda _E}x}}}}{{i!}}{{\left( {{\lambda _E}x} \right)}^i}},
\label{cdfgammase1}
\end{equation}
where ${\lambda _E} = \frac{{{m_E}}}{{{{\bar \gamma }_{SE}}}}$, ${\tau _E} = {m _E}{N _E}$, and
${\bar \gamma _{SE}}$ is the average SNR of $S$ - $E$ link.

Substituting (\ref{cdfgammaeq1}) and (\ref{pdfgammase1}) into (\ref{PoutL}), then utilizing \cite[(3.326.2)]{Gradshteyn2007}, \cite[(8)]{Lei2015CL}, and \cite[(21)]{VS1990}, we obtain
\begin{equation}
\begin{aligned}
P_{out, 1}^{\text{L}} &= \int_0^\infty  {{F_{\gamma _{eq, 1}}}\left( {\Theta \gamma } \right)} {f_{{\gamma _{SE,1}}}}\left( \gamma  \right)d\gamma \\
& = 1 - {\phi _R}\sum\limits_{S _R} {\sum\limits_{q = 0}^{B _R} {\sum\limits_{t = 0}^{q + {\tau _R} - 1} {{\Xi _1} \left( {{{\text{Z}}_0}\left( {t + {\tau _E} - 1} \right)!} \right.} } } \\
& - \left. {{B_D}\sum\limits_{h = 1}^\beta  {\sum\limits_{k = 0}^\infty  {{\phi _2}G_{r + 2,2r}^{r,r + 2}\left[ {{\upsilon _1}\left| {_{{K_2}}^{{K_1}}} \right.} \right]} } } \right),
\label{PoutL1}
\end{aligned}
\end{equation}
where
${\Xi _1} = \frac{{{\varphi _R}{\Theta ^t}\lambda _E^{{\tau _E}}}}{{\Gamma \left( {{\tau _E}} \right)\phi _1^{t + {\tau _E}}}}$,
$\phi _1 =  {\upsilon _R}\Theta  + {\lambda _E}$,
${\phi _2} = {H_1}\frac{{{r^{\frac{k}{2}}}{2^{t + {\tau _E} - 0.5}}}}{{{{\left( {2\pi } \right)}^{0.5r}}}}$,
${K_1} = \left[ {\Delta \left( {r,1} \right),\Delta \left( {2,1 - t - {\tau _E}} \right)} \right]$,
${K_2} = \left[ {\Delta \left( {r,\frac{{1 + k}}{2}} \right),\Delta \left( {r,0} \right)} \right]$,
$\Delta \left( {k,a} \right) = \left[\frac{a}{k},\frac{{a + 1}}{k}, \cdots ,\frac{{a + k - 1}}{k} \right]$, and
${\upsilon _1}{\text{ = }}\frac{{4\psi _1^r{\Theta ^2}}}{{{r^r}\phi _1^2}}$.

\begin{remark}
{One interesting conclusion can be found that the SOP under TASR scheme tends to be a constant, which means that the secrecy diversity order is zero. This is different from the results in \cite{Lei2018TCOM}, in which perfect CSI was assumed.}
\end{remark}

{
\textbf{Proof:}
The asymptotic SOP is expressed as
\begin{equation}
P_{out,1}^{{\rm{L}}\infty } = \int_0^\infty  {F_{{\gamma _{eq,1}}}^\infty \left( {\Theta \gamma } \right)} {f_{{\gamma _{SE,1}}}}\left( \gamma  \right)d\gamma,
\label{H232}
\end{equation}
where
\begin{equation}
\begin{aligned}
F_{{\gamma _{eq,1}}}^\infty \left( \gamma  \right)& = \mathop {\lim }\limits_{{{\bar \gamma }_{SR}} \to \infty } {F_{{{\hat \gamma }_{SR}}}}\left( \gamma  \right) + \mathop {\lim }\limits_{{{\bar \gamma }_{RD}} \to \infty } {F_{{{\hat \gamma }_{RD}}}}\left( \gamma  \right)\\
& - \mathop {\lim }\limits_{{{\bar \gamma }_{SR}} \to \infty } {F_{{{\hat \gamma }_{SR}}}}\left( \gamma  \right)\mathop {\lim }\limits_{{{\bar \gamma }_{RD}} \to \infty } {F_{{{\hat \gamma }_{RD}}}}\left( \gamma  \right).
\label{H232}
\end{aligned}
\end{equation}
Due to $\mathop {\lim }\limits_{{{\bar \gamma }_{RD}} \to \infty } {F_{{{\hat \gamma }_{RD}}}}\left( \gamma  \right) = 1 - {{\rm{Z}}_0}$, one can easily observe that $P_{out,1}^{{\rm{L}}\infty }$ tends to be a constant when ${\bar \gamma _{SR}} \to \infty$ and ${\bar \gamma _{RD}} \to \infty$. Thus, the secrecy diversity order is zero. Furthermore, this result also fits perfectly well with the SOP under TASE scheme for the same reason. In this work, we focus on the scenarios when ${\bar \gamma _{RD}} \to \infty$. The results for ${\bar \gamma _{SR}} \to \infty$ and ${\bar \gamma _{RD}} \to \infty$ can be easily obtained based on the results of this work.
}

Using the Meijer's $G$-function expansion given in \cite[(B.1)]{Zedini2014PJ}, we obtain the asymptotic SOP when ${\bar \gamma _{RD}} \to \infty$ as
\begin{equation}
\begin{aligned}
P_{out, 1}^{\text{L}, \infty }& = 1 + {\phi _R}\sum\limits_{S _R} {\sum\limits_{q = 0}^{B _R} {\sum\limits_{t = 0}^{q + {\tau _R} - 1} {{{\Xi _1} }} } }    \\
& \times \left( {B_D\sum\limits_{h = 1}^\beta  {\sum\limits_{k = 0}^\infty  {{\phi _2}} } {\Phi _{1}} - {{\text{Z}}_0}\left( {t + {\tau _E} - 1} \right)!} \right),
\label{PoutL1asy}
\end{aligned}
\end{equation}
where ${\Phi _{1}} = \sum\limits_{l = 1}^r {\frac{{\prod\limits_{j = 1,j \ne l}^r {\Gamma \left( {{K_{2,j}} - {K_{2,l}}} \right)} \prod\limits_{j = 1}^{r + 2} {\Gamma \left( {1 + {K_{2,l}} - {K_{1,j}}} \right)} }}{{\upsilon _{1} ^{ - {K_{2,l}}}\prod\limits_{j = r + 1}^{2r} {\Gamma \left( {1 + {K_{2,l}} - {K_{2,j}}} \right)} }}}  $.

\subsection{SOP with TAS based on $S$-$E$ link}
In the active eavesdropping scenario wherein the CSI of eavesdropping link is available at $S$\footnote{{More specifically, some users in the systems act as potential eavesdroppers  during certain instances while being the legitimate receivers  during other instances.
In other words, the source is simply aware of  the CSI of the eavesdroppers and is unaware of the location and identity of the eavesdroppers.
Furthermore, some potential eavesdroppers are multi-cast users and the messages transmitted by the source are confidential and meant for specific users only.}
},
the antenna can be selected based on the CSI of $S$-$E$ links, the TAS criterion in this scheme (TASE) can be expressed as
\begin{equation}
{b_2} = \arg \mathop {\min }\limits_{1 \le i \le {N_S}} \left( {\sum\limits_{j = 1}^{{N_E}} {{{\left| {{ h_{{S_i}{E_j}}}} \right|}^2}} } \right),
\label{bE}
\end{equation}
where $b_2$ signifies the selected antenna at $S$ based on $S$-$E$ link. Thus the actaul SNR at $E$ during data transmission can be written as
\begin{equation}
{{\hat \gamma }_{SE,2}} = \frac{{{P_S}}}{{{\sigma ^2}}}\sum\limits_{j = 1}^{{N_E}} {{{\left| {{{ \hat h}_{{S_{b_2}}{E_j}}}} \right|}^2}},
\label{gammasehat}
\end{equation}
where ${\hat h_{{S_{{b_2}}}{R_j}}}$ is the time-delayed version of ${h_{{S_{{b_2}}}{R_j}}}$.

Similarly, we have
\begin{equation}
{\hat h_{{S_i}{E_j}}} = {\rho _{SE}} {h_{{S_i}{E_j}}} + \sqrt {1 - {\rho _{SE} ^2}} \omega _2,
\label{hsehat}
\end{equation}
where ${\hat h_{{S_i}{E_j}}}$ and ${h_{{S_i}{E_j}}}$ are the outdated and accurate channel gain between the $i$-th antenna at $S$ and $j$-th antenna at $E$, respectively, $\omega _2$ represents a Nakagami-$m$ random variable with same variance as ${h_{{S_i}{E_j}}}$, and $0 < {\rho _{SE}} < 1$ is the correlation coefficient.

\begin{lemma}
\emph{The PDF and CDF of ${\hat \gamma _{SE,2}} $ are given as
\begin{equation}
{f_{{{\hat \gamma }_{SE,2}}}}\left( \gamma  \right) = {\phi _E}\sum\limits_{{S_E}} {\sum\limits_{q = 0}^{{B_E}} {{\Lambda _E}} } {\gamma ^{q + {\tau _E} - 1}}{e^{ - {\upsilon _E}\gamma }},
\label{pdfgammasehat}
\end{equation}
\begin{equation}
{F_{{{\hat \gamma }_{SE,2}}}}\left( \gamma \right) = 1 - {\phi _E}\sum\limits_{{S_E}} {\sum\limits_{q = 0}^{{B_E}} {\sum\limits_{t = 0}^{q + {\tau _E} - 1} {{\varphi _E}{\gamma^t}{e^{ - {\upsilon _E}\gamma}}} } },
\label{cdfgammasehat}
\end{equation}
respectively, where
${\phi _E} = \frac{{{N_S}\lambda _E^{{\tau _E} + 1}\rho _{SE}^{1 - {\tau _E}}}}{{\Gamma \left( {{\tau _E}} \right)\left( {1 - \rho _{SE}^2} \right)}}$,
${S _E} = \left\{ {\left( {{n_1}, \cdots, {n_{{\tau _E} }}} \right) \in \mathbb{N}\left| {\sum\limits_{p = 1}^{{\tau _E}} {{n_p} = {N_S} - 1} } \right.} \right\}$,
${\beta _E} = \frac{{ {{\rho _{SE}}} }}{{1 - \rho _{SE}^2}} $,
${\upsilon _E} =  {\frac{{{\lambda _E}}}{{1 - \rho _{SE}^2}} - \frac{{{{\left( {{\lambda _E}\beta _E} \right)}^2}}}{\alpha _E}} $,
${\varphi _E} = \frac{{{\Lambda _E}\left( {q + {\tau _E} - 1} \right)!{\upsilon _E^{t - q - {\tau _E}}}}}{{t!}}$,
${\alpha _E} = {\lambda _E}\left( {\frac{{{\rho _{SE}^2}}}{{1 - \rho _{SE}^2}} + {C_E} + 1} \right) $,
${A_E} = \left( {\frac{{{N_S} - 1}}{{\prod\limits_{q = 1}^{{\tau _E}} {{n_q}} }}} \right)\prod\limits_{p = 1}^{{\tau _E}} {{{\left( {\frac{{\lambda _E^{p - 1}}}{{\left( {p - 1} \right)!}}} \right)}^{{n_p}}}} $,
${B _E} = \sum\limits_{p = 1}^{{\tau _E}} {{n_p}\left( {p - 1} \right)} $,
${C_E} = \sum\limits_{p = 1}^{{\tau _E}} {{n_p}} $, and
${\Lambda _E} = \frac{{{A _E}{B _E}!{{\left( {{\lambda _E}\beta _E} \right)}^{2q + {\tau _E} - 1}}\left( {{B _E} + {\tau _E} - 1} \right)!}}{{q!{\alpha _E^{{B _E} + {\tau _E} + q}}\left( {{B _E} - q} \right)!\left( {{\tau _E} + q - 1} \right)!}}$.}
\end{lemma}

{\it {Proof :}} See Appendix C. \\

Similarly, TAS based on the CSI of $S$-$E$ link signifies that a random transmit antenna would be selected for $R$. Thus the SNR at $R$ under this scheme can be expressed as
\begin{equation}
{\gamma _{SR, 2}} = \frac{{{P_S}}}{{{\sigma ^2}}}\sum\limits_{j = 1}^{{N_R}} {{{\left| {{h_{{S_{b_2}}{R_j}}}} \right|}^2}}.
\end{equation}
The PDF and CDF of ${\gamma _{SR, 2}} $ can be easily obtained as
\begin{equation}
{f_{{\gamma _{SR, 2}}}}\left( x \right) = \frac{{{\lambda _R^{{\tau _R}}}}}{{\Gamma \left( {{\tau _R}} \right)}}{e^{ - {\lambda _R}x}}{x^{{\tau _R} - 1}},
\label{pdfgammasr2}
\end{equation}
\begin{equation}
{F_{{\gamma _{SR, 2}}}}\left( x \right) = 1 - \sum\limits_{i = 0}^{{\tau _R} - 1} {\frac{{{e^{ - {\lambda _R}x}}}}{{i!}}{{\left( {{\lambda _R}x} \right)}^i}}.
\label{cdfgammasr2}
\end{equation}
Finally, we obtain the CDF of $\gamma _{eq}$ in this case as
\begin{equation}
\begin{aligned}
{F_{{\gamma _{eq,2}}}}\left( \gamma  \right) &= {F_{{\gamma _{SR,2}}}}\left( \gamma  \right) + {F_{{{\hat \gamma }_{RD}}}}\left( \gamma  \right) - {F_{{\gamma _{SR,2}}}}\left( \gamma  \right){F_{{{\hat \gamma }_{RD}}}}\left( \gamma  \right)\\
& = 1 + \sum\limits_{i = 0}^{{\tau _R} - 1} {\frac{{{e^{ - {\lambda _R}\gamma }}}}{{i!}}{{\left( {{\lambda _R}\gamma } \right)}^i}} \\
& \times \left( {{B_D}\sum\limits_{h = 1}^\beta  {\sum\limits_{k = 0}^\infty  {{H_1}} } G_{1,2}^{1,1}\left[ {{{\psi _1}}{\gamma ^{\frac{2}{r}}}\left| {_{\frac{{1 + k}}{2},0}^1} \right.} \right] - {Z_0}} \right).
\label{cdfgammaeq2}
\end{aligned}
\end{equation}

%
\begin{remark}
Based on (\ref{gammaeq}), one can observe that when ${\hat \gamma _{SR}} > {\tilde \gamma _{RD}}$, which means the FSO link is the bottleneck of the equivalent SNR at the destination, the TAS based on $S$-$E$ link will be the optimal TAS scheme because EST in this case is independent of $S$-$R$ links.
\end{remark}

Substituting (\ref{pdfgammasehat}) and (\ref{cdfgammaeq2}) into (\ref{PoutL}) and after some algebraic manipulations, we obtain
\begin{equation}
\begin{aligned}
P_{out,2}^\text{L} &= \int_0^\infty  {{F_{{\gamma _{eq,2}}}}\left( {\Theta \gamma } \right)} {f_{{\hat \gamma _{SE,2}}}}\left( \gamma  \right)d\gamma \\
& = 1 - {\phi _E}\sum\limits_{{S_E}} {\sum\limits_{q = 0}^{{B_E}} {\sum\limits_{i = 0}^{{\tau _R} - 1} {{\Lambda _E}{{\Xi _2} }\left( {{{{\text{Z}}_0}}\left( {i + q + {\tau _E} - 1} \right)!} \right.} } } \\
& \left. { - {B_D}\sum\limits_{h = 1}^\beta  {\sum\limits_{k = 0}^\infty  {\phi _4} } G_{r + 2,2r}^{r,r + 2}\left[ {{\upsilon _2}\left| {_{{K_2}}^{{K_3}}} \right.} \right]} \right),
\label{PoutL2}
\end{aligned}
\end{equation}
where
${\Xi _2 } = \frac{{{{\left( {\Theta {\lambda _R}} \right)}^i}}}{{\phi _3^{i + q + {\tau _E}}i!}}$,
${\upsilon _2} = \frac{{4\psi _1^r\Theta^2}}{{{r^r}\phi _3^2}}$,
${\phi _3} = \Theta {\lambda _R} + {\upsilon _E}$,
$\phi _4 = \frac{{{H_1}{r^{\frac{k}{2}}}{2^{i + q + {\tau _E} - 0.5}}}}{{{{\left( {2\pi } \right)}^{0.5r}}}}$, and
${K_3} = \left[ {\Delta \left( {r,1} \right),\Delta \left( {2,1 - i - q - {\tau _E}} \right)} \right]$.

Similar to (\ref{PoutL1asy}), we obtain the asymptotic SOP for this scenario as
\begin{equation}
\begin{aligned}
P_{out, 2}^{\text{L}, \infty } & = 1 + {\phi _E}\sum\limits_{{S_E}} {\sum\limits_{q = 0}^{{B_E}} {\sum\limits_{i = 0}^{{\tau _R} - 1} {{\Lambda _E}{{\Xi _2} }} } } \\
& \times \left( {{B_D}\sum\limits_{h = 1}^\beta  {\sum\limits_{k = 0}^\infty  {\phi _4} } {\Phi _2} - {{{\text{Z}}_0}}\left( {i + q + {\tau _E} - 1} \right)!} \right),
\label{PoutL2asy}
\end{aligned}
\end{equation}
where ${\Phi _2} = \sum\limits_{l = 1}^r {\frac{{\prod\limits_{j = 1,j \ne l}^r {\Gamma \left( {{K_{2,j}} - {K_{2,l}}} \right)} \prod\limits_{j = 1}^{r + 2} {\Gamma \left( {1 + {K_{2,l}} - {K_{3,j}}} \right)} }}{{\upsilon _2^{ - {K_{2,l}}}\prod\limits_{j = r + 1}^{2r} {\Gamma \left( {1 + {K_{2,l}} - {K_{2,j}}} \right)} }}} $.

There is another interesting question: when ${\hat \gamma _{SR}} < {\tilde \gamma _{RD}}$, between the TAS scheme based on $S$-$R$ links or $S$-$E$ links, which one can obtain better secrecy performance? The answer depends on the relationship between the SNRs of $S$-$R$ and $S$-$E$ links.

\subsection{A new adaptive TAS Scheme}

An interesting problem is ``Which scheme can obtain better secrecy performance between TASR and TASE?" It depends on the channel quality of the RF and FSO links. The reasons are given as follow.

1) When ${\bar \gamma _{SR}} > {\bar \gamma _{RD}}$, the FSO link is the bottleneck of the equivalent SNR at $D$. In this case, the SOP depends on the $S$-$E$ and FSO links. Thus, TASE will obtain better secrecy performance than that of TASR.

\begin{table*}[!htp]
\centering
\caption{{\textbf{Comparison of TAS Scheme}}}
\begin{threeparttable}
\begin{tabular}{c | c|c |c | c }
\hline
\textbf{Scheme} & \textbf{CSI Requirement } & \textbf{Criterion} & \textbf{Expression for SOP} & \textbf{Condition} \\
\hline
OTAS &  S-R, S-E, and R-D                  & Eq. (\ref{bOTAS})  & -                  & -                \\
\hline
TASR & selected index of S-R               & Eq. (\ref{bR})     & Eq. (\ref{PoutL1}) & $\mho $\tnote{a} \\
\hline
TASE & selected index of S-E               & Eq. (\ref{bE})     & Eq. (\ref{PoutL2}) &  $1 - \mho$       \\
\hline
ATAS &selected index of S-R and S-E        & Eq. (\ref{b3})     & Eq. (\ref{sopATAS})& -                 \\
\hline
\end{tabular}
\begin{tablenotes}
  \item[a] {$ \mho  = \left\{ {{{\bar \gamma }_{SE}} < {{\bar \gamma }_{SR}} < {{\bar \gamma }_{RD}}} \right\}$}
\end{tablenotes}
 \end{threeparttable}
\label{table}
\end{table*}

2) Otherwise, in the case that $S$-$R$ link is the bottleneck of the equivalent SNR at $D$, the SOP depends on the $S$-$R$ and $S$-$E$ links. When the CSI of $S$-$E$ is unavailable, we can not help but select antenna based on $S$-$R$ link. The antenna is selected based on the link that has larger average SNR since large SNR has a strong influence on the SOP of mixed RF-FSO systems.\footnote{Actually, when the eavesdropper's CSI is available, the source node can transmit artificial noise to enhance the secrecy performance of the mixed RF-FSO systems, which will be conducted in our future work based on the results of this paper.}

To obtain a LOS transmission, $R$ (the sender of the FSO link) must be located at some place with high altitude, such as the building roof. In other words, the location of $R$ is limited while designing a mixed RF-FSO system. Thus we propose an adaptive TAS (ATAS) scheme, the criterion of which is expressed as
\begin{equation}
{b_3} = \left\{ {\begin{array}{*{20}{c}}
{{b_2},}&{{{\bar \gamma }_{SR}} > {{\bar \gamma }_{RD}}}\\
{{b_2},}&{\left( {{{\bar \gamma }_{SR}} < {{\bar \gamma }_{RD}}} \right) \cup \left( {{{\bar \gamma }_{SR}} < {{\bar \gamma }_{SE}}} \right)}\\
{{b_1},}&{\left( {{{\bar \gamma }_{SR}} < {{\bar \gamma }_{RD}}} \right) \cup \left( {{{\bar \gamma }_{SR}} > {{\bar \gamma }_{SE}}} \right)}
\end{array}} \right.
\label{b3}
\end{equation}

The ATAS scheme is explained as follows. When the relay is located in those places that makes
${{{\bar \gamma }_{SR}} > {{\bar \gamma }_{RD}}}$, $R$-$D$ link becomes the bottleneck of the equivalent SNR at $D$.
Under this scenario, the secrecy performance of the considered system depends on $S$-$E$ and $R$-$D$ links.
Thus, we select antenna based on $S$-$E$ link (defined by (\ref{bE})) where the CSI of $E$ is known at $S$.
On the other hand, when $R$ is located on those places that makes ${{{\bar \gamma }_{SR}} < {{\bar \gamma }_{RD}}}$, $S$-$R$ link becomes bottleneck of the equivalent SNR at $D$.
Under such a scenario, the secrecy performance of the considered system depends on the first hop.
Thus, we select antenna based on $S$-$R$ link (defined by (\ref{bR})) when $S$ does not has the CSI of $E$.
Moreover, when $S$ has the CSI of $E$ and ${{{\bar \gamma }_{SR}} < {{\bar \gamma }_{RD}}}$, we select the antenna based on the CSI of $S$-$E$ links when ${{{\bar \gamma }_{SR}} < {{\bar \gamma }_{SE}}}$.
Otherwise, we select the antenna simply based on the CSI of $S$-$R$ links when ${{{\bar \gamma }_{SR}} > {{\bar \gamma }_{SE}}}$.

{Obviously, one can obtain the ATAS scheme simple based on the average SNR of RF links and FSO link, which are usually easy to obtain.} But the results are rough and not always the best when the difference between the RF link and FSO link is small (see Fig. \ref{fig02}).

{Based on (\ref{b3}), one can obtain the bound of SOP with ATAS scheme as
\begin{equation}
P_{out,{\rm{ATAS}}}^{\rm{L}} = \left\{ {\begin{array}{*{20}{c}}
{P_{out,1}^{\rm{L}},}&{\left( {{{\bar \gamma }_{SR}} < {{\bar \gamma }_{RD}}} \right) \cup \left( {{{\bar \gamma }_{SR}} > {{\bar \gamma }_{SE}}} \right)}\\
{P_{out,2}^{\rm{L}},}&{{\rm{others}}}
\end{array}} \right..
\label{sopATAS}
\end{equation}}

\section{Comparison and Discussion}
\label{sec:Discussion}
{To identify the strengths and weaknesses of each proposed scheme, Table \ref{table} summarizes a more detailed secrecy outage performance comparison for all the proposed schemes. In general, realization of all the TAS schemes depends heavily on the available CSI at the source.
The more CSI is available, the better secrecy outage performance will be obtained.
The OTAS can obtain the `best' secrecy outage performance since all the links are being utilized, which has been testified in Figs. \ref{fig02} - \ref{fig05}. However, a closed-form expression for the SOP of the OTAS can not be obtained, as shared in Section III.A. It must be noted that the quality of the CSI also has significant influence on the secrecy performance with all the TAS schemes, which has been testified in Figs. \ref{fig06} - \ref{fig11}.
When the CSI of eavesdropping links is not available at the source node, the TASR scheme is the unique choice.
Otherwise, either TASR or TASE schemes can be utilized.
One can realize the SOP under the TASR outperforms that under the TASE when condition $\mho$ is satisfied since the $S$-$R$ link is the bottleneck of secrecy performance for the mixed RF-FSO systems in this scenario. And this point will be demonstrated in Section V. In other words, for those scenarios where the CSI of eavesdropping links is available but condition $\mho$ is not satisfied, the TASE scheme is a better choice due to its simplicity. The ATAS scheme obtains the second best secrecy performance as it combines the advantages of both the TASR and TASE schemes. It must be noted that ATAS is simply based on the average SNR of RF links and FSO link, which are very often easy to obtain. Thus, the results are rough and not always the best when the difference between the RF and FSO links is small.}


\section{Numerical Results and Discussions}
\label{sec:RESULTS}

In this part, we present some representative Monte-Carlo simulation results to explain the behaviors of EST with respect to different TAS schemes and the parameters of the considered system.
We set $\rho _{SR} = \rho _{SE} = \rho _{\text{RF}}$,
${m_R} = {m_E} = m$, ${R_s} = 0.01\,{\text{bit/s/Hz}}$.
{All the parameters of FSO link utilized here are inspired from \cite{Ansari2016TWC} and are set as
the FSO link distance $L = 1$ km,
the wavelength $ {\lambda _{{\rm{FSO}}}} = 785$ nm,
the refraction structure parameter $C_n^2 = 1.2 \times {10^{ - 13}}\,{m^{ - \frac{2}{3}}}$,
the distance dependent path loss in the clear weather ${I_l} = 0.9$ / km,
the fraction of the collected optical power ${A_0} = 1$,
the average power of the LOS component $\Omega  = 1.3265$,
the average power of the total scatter components ${b_0} = 0.1079$,
the amount of scattering power coupled to the LOS component ${\rho _0} = 0.596$,
the difference between phases of the LOS and the coupled-to-LOS  ${\varphi _A} - {\varphi _B} = \frac{\pi }{2}$,
and
the scintillation parameters $\alpha  = 2.296$, and $\beta  = 2$.}
{In calculation of the infinite summation terms in the analytical expressions, we truncate the infinite terms into 80 terms.}
In all the figures,
`Sim' denotes the simulation results.
One can observe that the simulation results match the numerical results very well, which testified that the bound obtained in (\ref{PoutL}) is tight and {the results in this work are convergent}.

\begin{figure}[!t]
\centering
\includegraphics[width = 3.05 in]{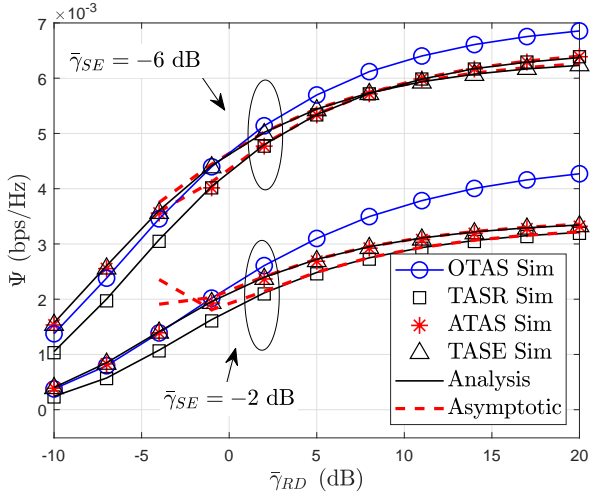}
\caption{EST for varying ${\bar \gamma _{SE}}$ with $N_S = 5$, $N_R = N_E = 2$, $\alpha = 2.296$, $\beta = 2$, $\xi = 6.7$, $r  = 2$, ${\rho _{\text{RF}} = 0.85}$, ${\rho _{\text{FSO}} = 0.5}$, $m = 2$, and ${\bar \gamma _{SR}} = -4\,{\text{dB}}$.}
\label{fig02}
\end{figure}

\begin{figure}[!t]
\centering
\includegraphics[width = 3.05 in]{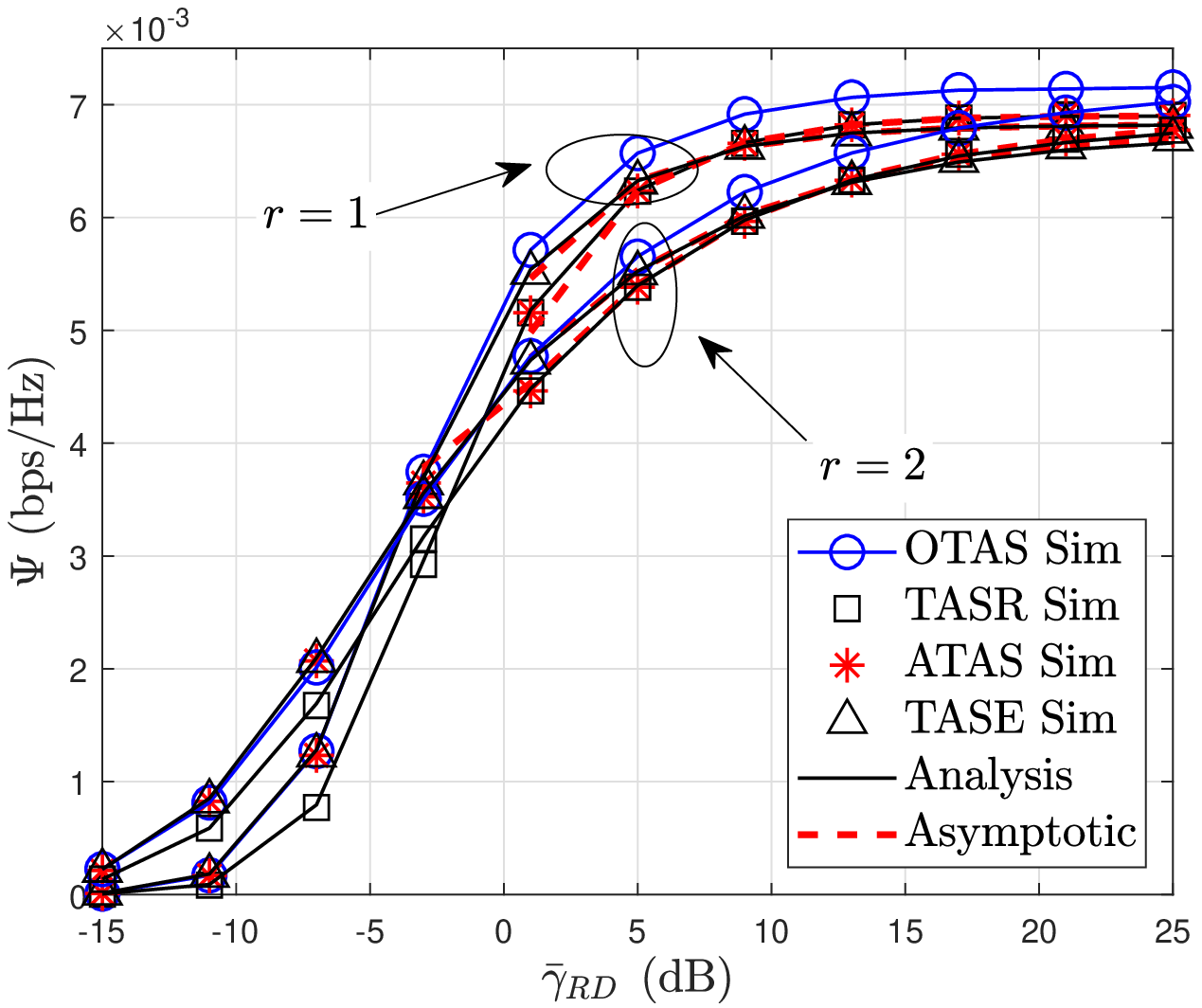}
\caption{EST for varying $r$ with $N_S = 5$, $N_R = N_E = 2$, $\alpha = 2.296$, $\beta = 2$, $\xi  = 6.7$, ${\rho _{\text{RF}} = 0.7}$, ${\rho _{\text{FSO}} = 0.5}$, $m = 2$, ${\bar \gamma _{SR}} = -1,{\text{dB}}$, and ${\bar \gamma _{SE}} = -5\,{\text{dB}}$.}
\label{fig03}
\end{figure}

\begin{figure}[!t]
\centering
\includegraphics[width = 3.05 in]{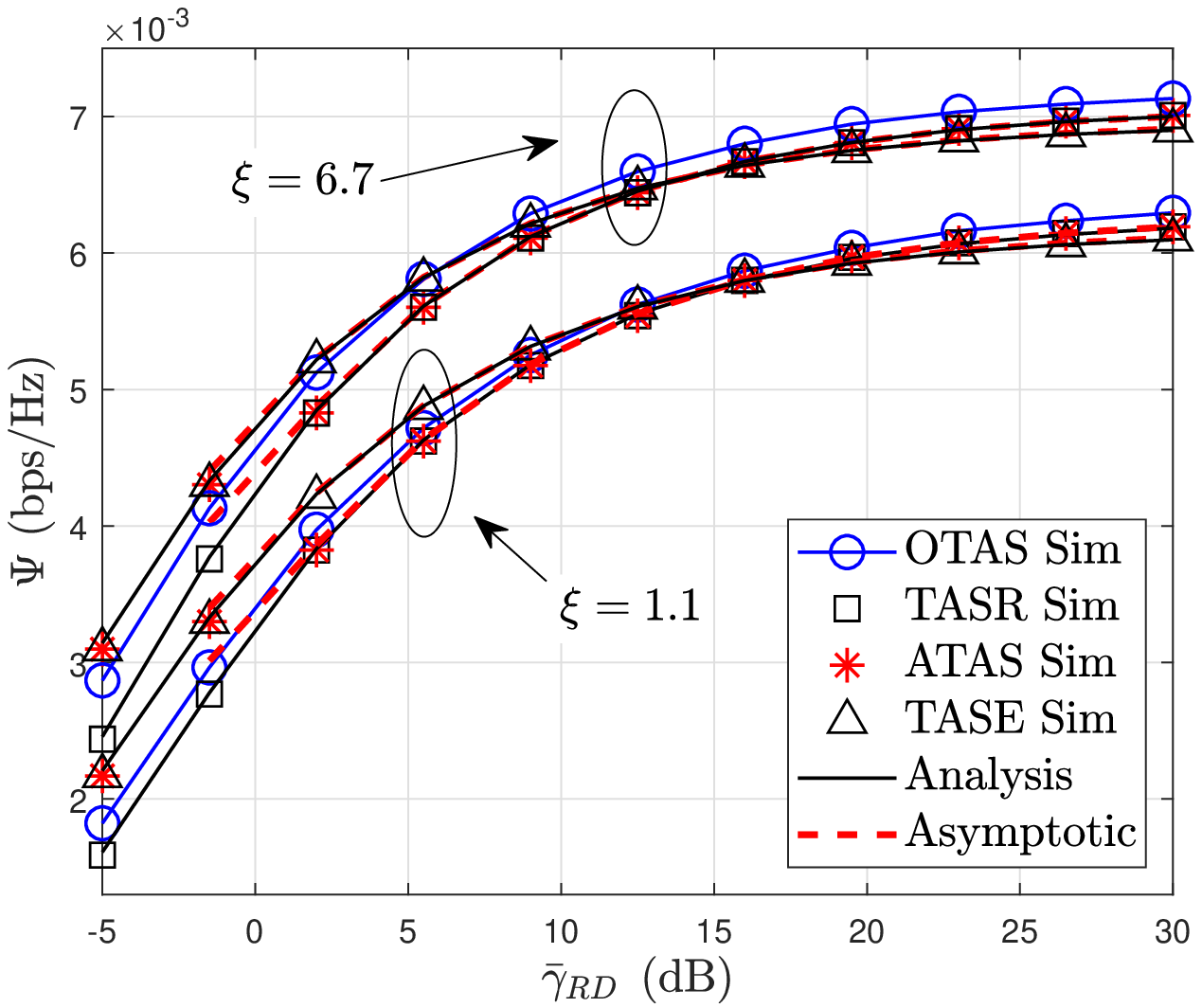}
\caption{EST for varying $\xi$ with $N_S = 5$, $N_R = N_E = 2$, $\alpha = 2.296$, $\beta = 2$, $r  = 2$, ${\rho _{\text{RF}} = 0.85}$, ${\rho _{\text{FSO}} = 0.5}$, $m = 2$, ${\bar \gamma _{SR}} = -1\,{\text{dB}}$, and ${\bar \gamma _{SE}} = -5\,{\text{dB}}$.}
\label{fig04}
\end{figure}

\begin{figure}[!t]
\centering
\includegraphics[width = 3.05 in]{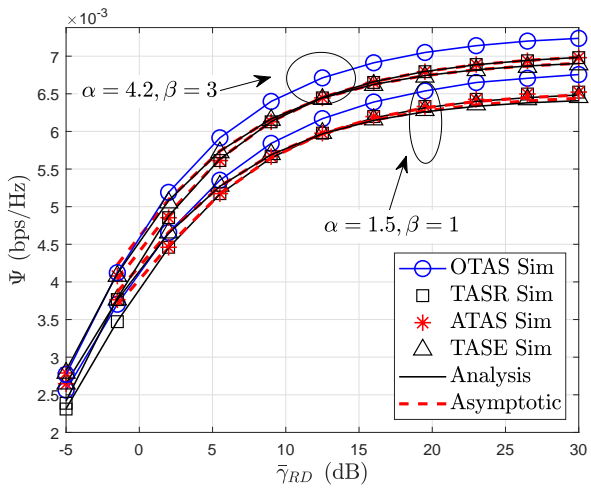}
\caption{EST for varying ($\alpha, \beta$) with $N_S = 4$, $N_R = N_E = 2$, $\xi = 6.7$, $r  = 2$, ${\rho _{\text{RF}} = 0.7}$, ${\rho _{\text{FSO}} = 0.5}$, $m = 2$, ${\bar \gamma _{SR}} = -1\,{\text{dB}}$, and ${\bar \gamma _{SE}} =-5\,{\text{dB}}$.}
\label{fig05}
\end{figure}

Figs. \ref{fig02} - \ref{fig05} show the EST with different TAS schemes for various ${\bar \gamma _{SE}}$, $r$, $\xi$, and $\left( {\alpha ,\beta } \right)$, respectively. The EST increases significantly with increasing ${\bar \gamma _{RD}}$ and there is a ceiling since the equivalent SNRs at $D$ will be equal to the SNR of $S$-$R$ link.
Furthermore, one can observe that the EST with TASE is approximate to that of OTAS, and is better than that of TASR in lower-${\bar \gamma _{RD}}$ region.
This is because in these areas $S$-$R$ links have little effect on the EST, which depend on the $R$-$D$ and $S$-$E$ links. In the higher-${\bar \gamma _{RD}}$ region, TASR scheme can obtain better secrecy performance compared with TASE scheme when ${{{\bar \gamma }_{SR}} > {{\bar \gamma }_{SE}}}$, which can be found from Fig. \ref{fig02}.
Vice versa, in those scenarios when ${{{\bar \gamma }_{SR}} < {{\bar \gamma }_{SE}}}$, the EST of TASE outperforms that of TASR since $S$-$E$ link is the major factor.

From Figs. \ref{fig03}, \ref{fig04}, and \ref{fig05}, it can be observed that the EST with $r = 1$ or $\xi = 6.7$ or ($\alpha = 4.3$, $\beta = 3$) outperforms that with $r = 2$ or $\xi = 1.1$ or ($\alpha = 1.5$, $\beta = 1$) since smaller SNR can be obtained for $R$-$D$ link in the former cases relative to the latter ones. These latter scenarios signify IM/DD detect technology, larger pointing error, and stronger atmospheric turbulence conditions, respectively.

\begin{figure}[!t]
\centering
\includegraphics[width = 3.05 in]{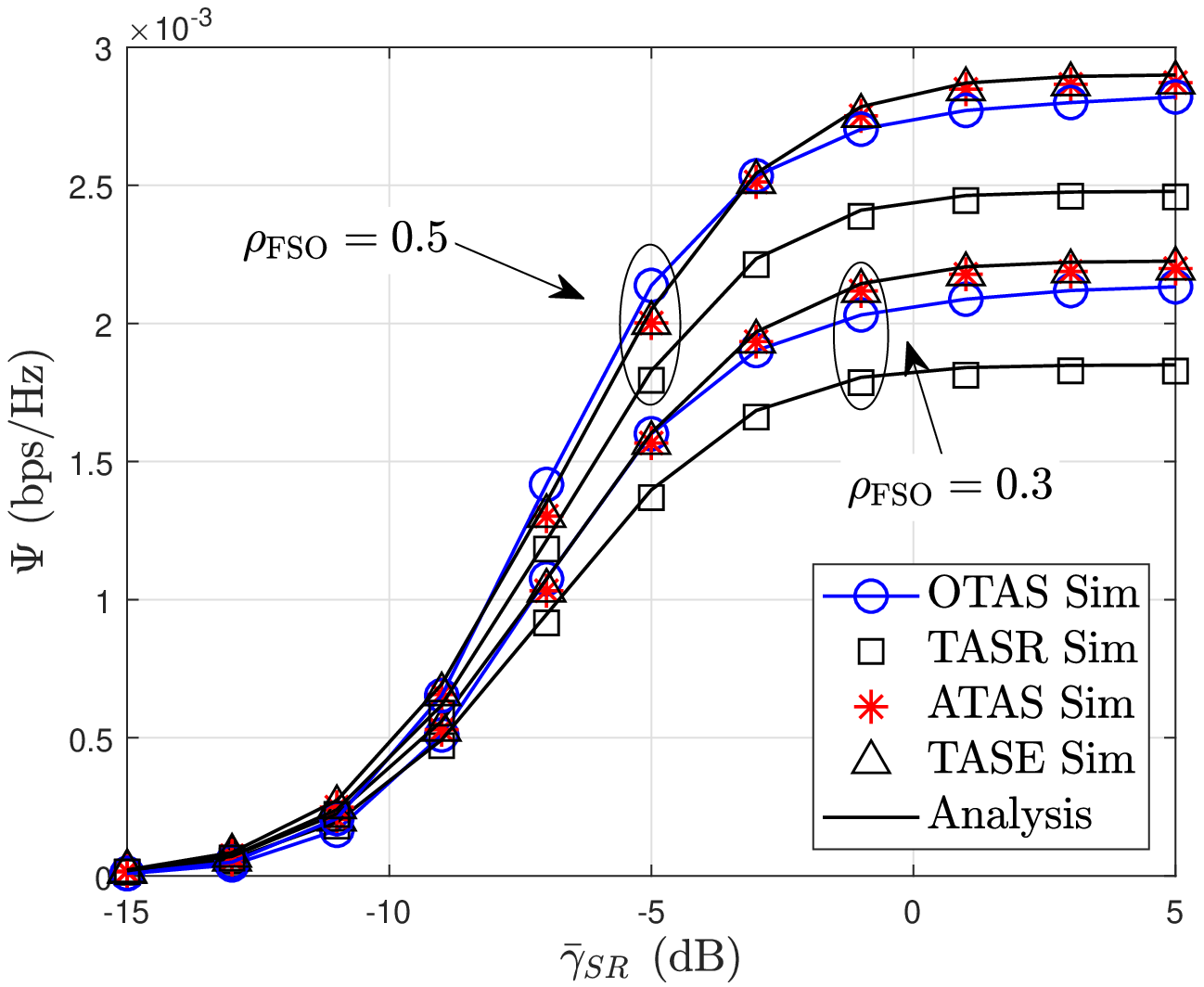}
\caption{EST for varying ${\rho _{\text{FSO}}}$ with $N_S = 4$, $N_R = N_E = 2$, $\alpha = 2.296$, $\beta = 2$, $\xi = 6.7$, $r = 2$, ${\rho _{\text{RF}} = 0.7}$, $m = 2$, ${\bar \gamma _{RD}} = -5\,{\text{dB}}$, and ${\bar \gamma _{SE}} = -5\,{\text{dB}}$.}
\label{fig06}
\end{figure}

\begin{figure}[!t]
\centering
\includegraphics[width = 3.05 in]{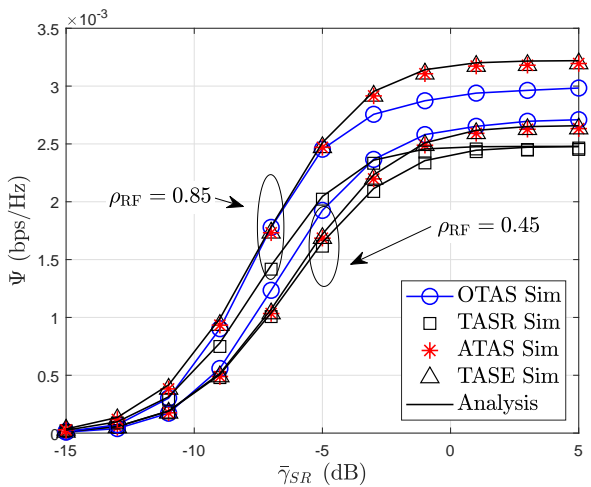}
\caption{EST for varying ${\rho _{\text{RF}}}$ with $N_S = 5$, $N_R = N_E = 2$, $\alpha = 2.296$, $\beta = 2$, $r  = 2$, $\xi = 6.7$, $m = 2$, ${\rho _{\text{FSO}} = 0.5}$, ${\bar \gamma _{RD}} = -5\,{\text{dB}}$, and ${\bar \gamma _{SE}} = -5\,{\text{dB}}$.}
\label{fig07}
\end{figure}

\begin{figure}[!t]
\centering
\includegraphics[width = 3.05 in]{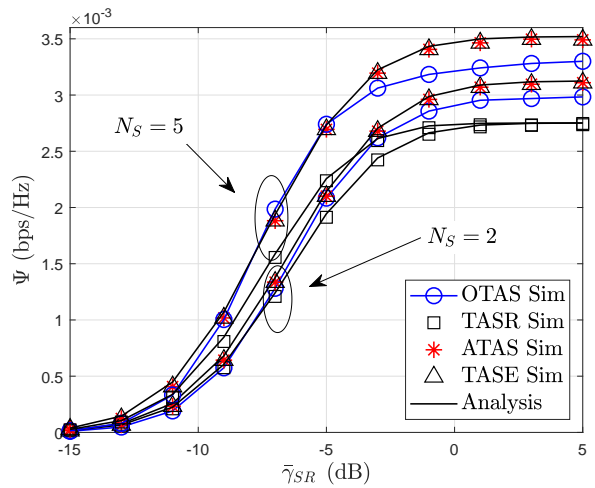}
\caption{EST for varying $N_S$ with $N_R = 2$, $N_E = 2$, $\alpha = 2.296$, $\beta = 2$, $\xi = 6.7$, $r  = 2$, ${\rho _{\text{RF}}= 0.85}$, ${\rho _{\text{FSO}} = 0.6}$, $m = 2$, ${\bar \gamma _{RD}} = -5\,{\text{dB}}$, and ${\bar \gamma _{SE}} = -5\,{\text{dB}}$.}
\label{fig08}
\end{figure}

\begin{figure}[!t]
\centering
\includegraphics[width = 3.05 in]{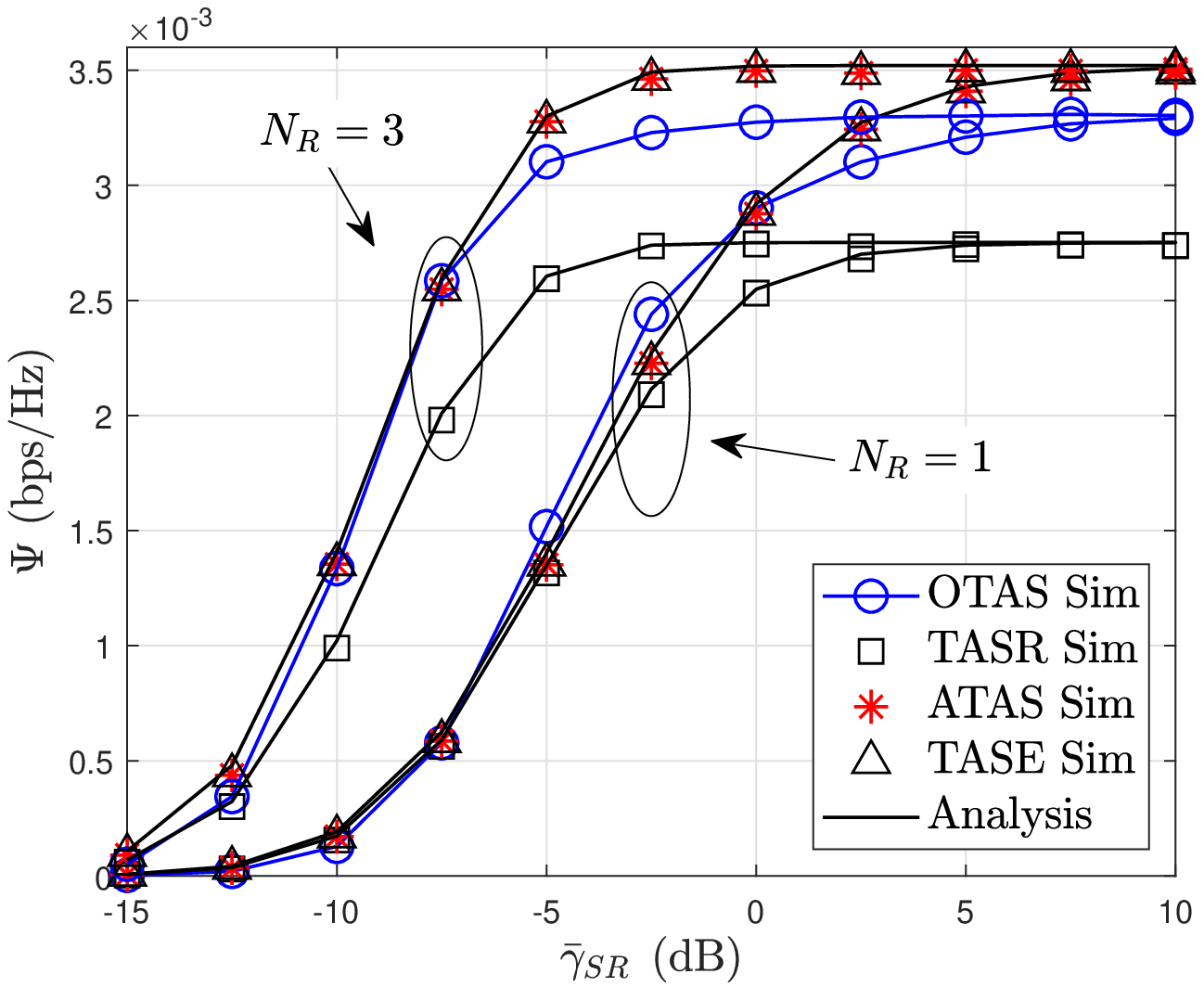}
\caption{EST for varying $N_R$ with $N_S = 5$, $N_E = 2$, $\alpha = 2.296$, $\beta = 2$, $\xi = 6.7$, $r  = 2$, ${\rho _{\text{RF}}= 0.85}$, ${\rho _{\text{FSO}} = 0.6}$, $m = 2$, ${\bar \gamma _{RD}} = -5\,{\text{dB}}$, and ${\bar \gamma _{SE}} = -5\,{\text{dB}}$.}
\label{fig09}
\end{figure}

\begin{figure}[!t]
\centering
\includegraphics[width = 3.05 in]{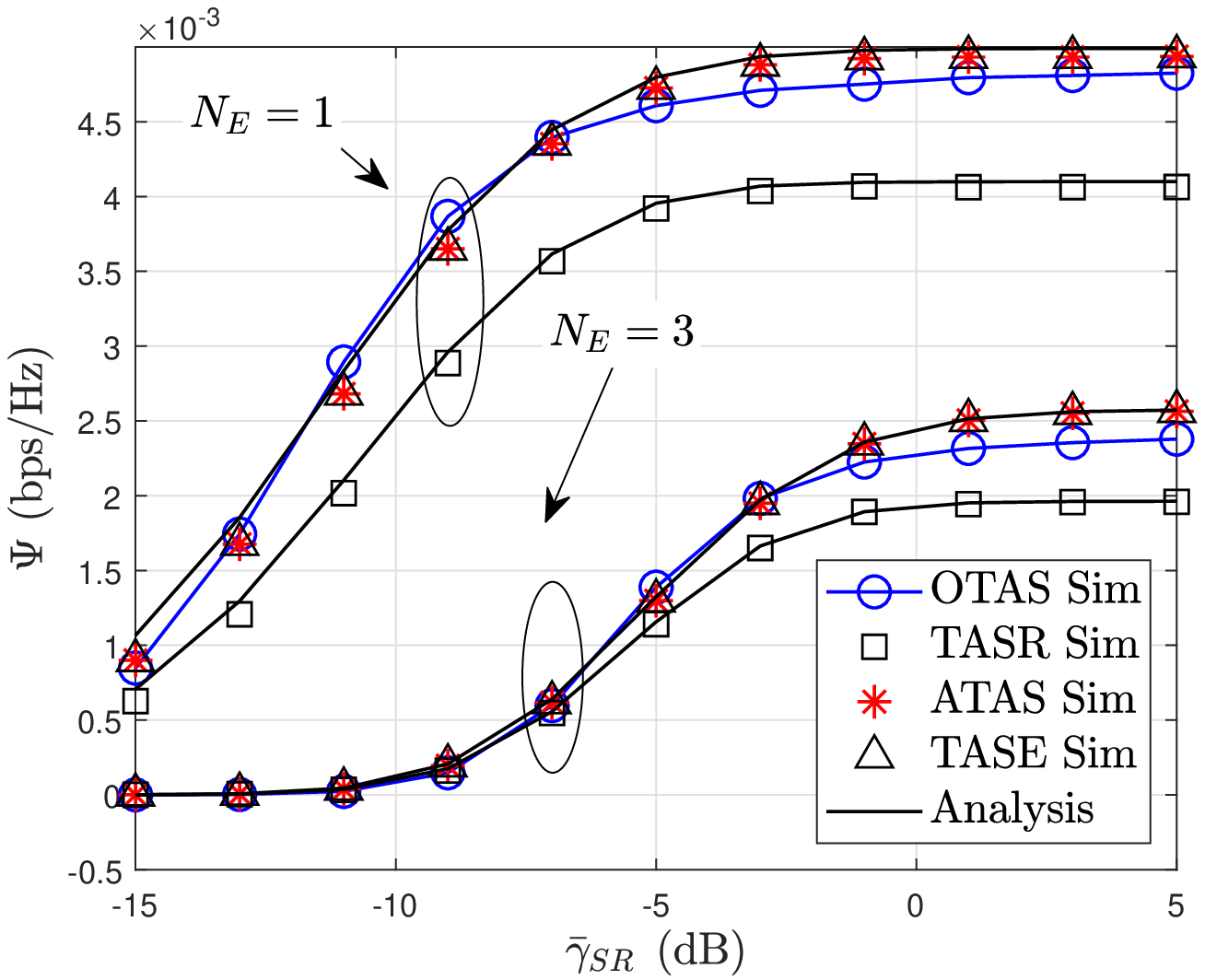}
\caption{EST for varying $N_E$ with $N_S = 5$, $N_R = 2$, $\alpha = 2.296$, $\beta = 2$, $\xi = 6.7$, $r  = 2$, ${\rho _{\text{RF}} = 0.85}$, ${\rho _{\text{FSO}} = 0.6}$, $m = 2$, ${\bar \gamma _{RD}} = -5\,{\text{dB}}$, and ${\bar \gamma _{SE}} = -5\,{\text{dB}}$.}
\label{fig10}
\end{figure}

\begin{figure}[!t]
\centering
\includegraphics[width = 3.05 in]{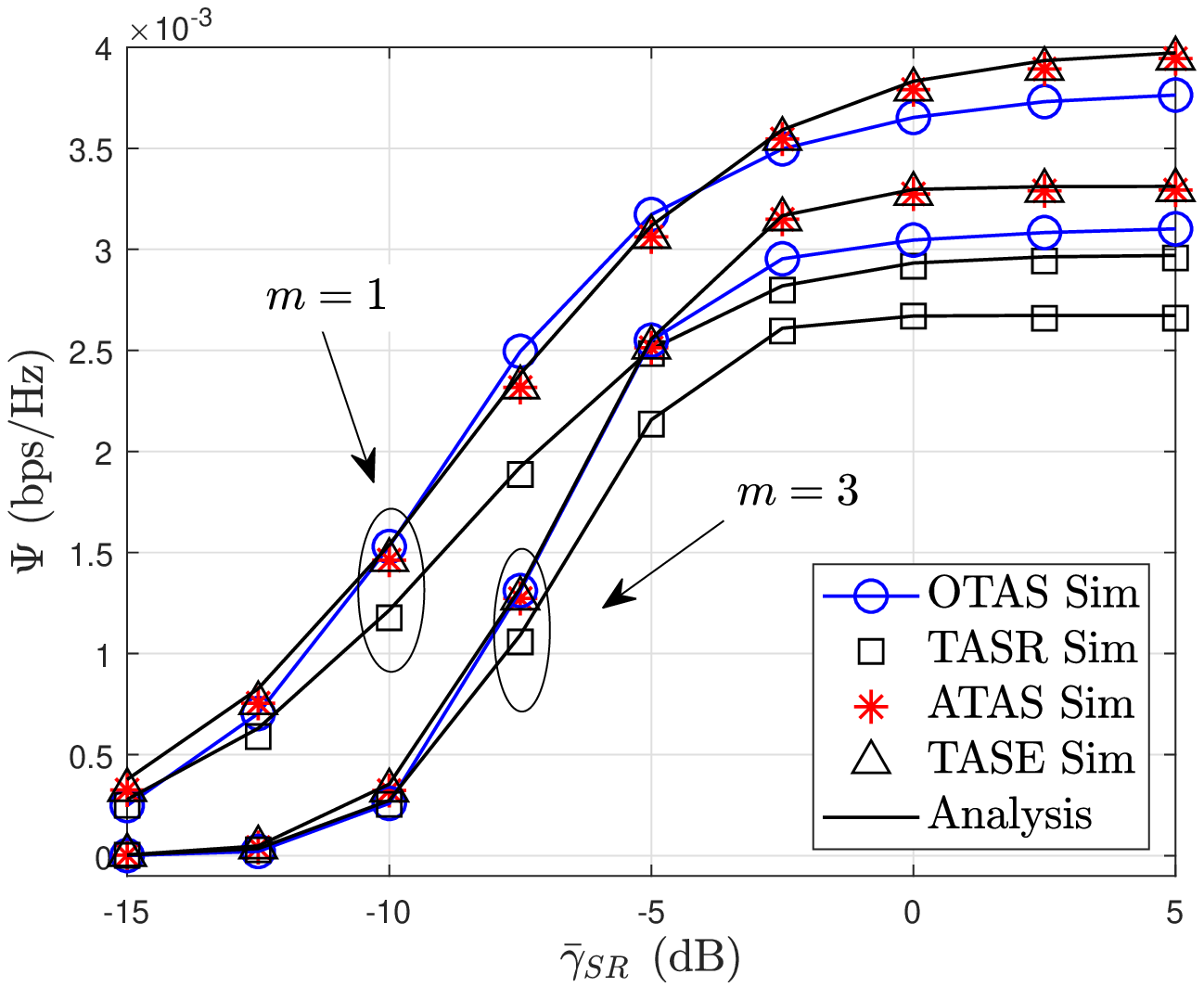}
\caption{EST for varying $m$, $N_S = 5$, $N_R = N_E = 2$, $\alpha = 2.296$, $\beta = 2$, $\xi = 6.7$, $r  = 2$, ${\rho _{\text{RF}} = 0.85}$, ${\rho _{\text{FSO}} = 0.6}$, ${\bar \gamma _{RD}} = -5\,{\text{dB}}$, and ${\bar \gamma _{SE}} = -5\,{\text{dB}}$.}
\label{fig11}
\end{figure}

Figs. \ref{fig06} and \ref{fig07} demonstrate that the EST with larger correlation coefficients outperform the ones with smaller correlation coefficient because the smaller correlation coefficients signify severely outdated effect (for RF links) or larger channel estimation errors (for FSO links). One can find that the EST with TASE outperforms those of OTAS and TASR in the higher-${\bar \gamma _{SR}}$ region. This is because the secrecy outage performance in this region depend on $S$-$E$ and $R$-$D$ links and ${\rho _{\text{RF}}}$ does not influence the EST.
{Moreover, the OTAS does not obtain the real optimal performance due to the outdated effect and channel estimation errors. The same results can also be obtained from Figs. \ref{fig08}-\ref{fig11}.}

Figs. \ref{fig08} -  \ref{fig11} demonstrate the EST with different TAS schemes for various $N_S$, $N_R$, $N_E$, and $m$, respectively.
One can draw conclusion from Fig. \ref{fig08} that increasing $N_S$ is not an effective method to enhance the secrecy performance of the considered system.
As observed from Figs. \ref{fig09} and \ref{fig10}, we can observe that the effect of the antenna number of $E$ on the EST is greater than that of the antenna number of $R$, especially for the EST with TASE and OTAS.
From Fig. \ref{fig11}, we can deduce that in the lower-${\bar \gamma _{SR}}$ region, the EST with smaller $m$ is better than that with larger $m$, which is consistent with the results in \cite{YangN2013TCOM}. In this region, $S$-$R$ link is the bottleneck for the equivalent SNR at $D$ and the system behaves synonymous to a MIMO RF Wyner model, which was considered with perfect CSI in \cite{YangN2013TCOM}.


Furthermore, relative to the results in \cite{Lei2017PJ}, one can observe that the secrecy performance of the considered system deteriorates due to imperfect CSI, which is consistent with the results in \cite{Lei2018PJ}.
\section{Conclusion}
\label{sec:Conclusion}

Four different TAS schemes were proposed to improve the secrecy performance of mixed RF-FSO systems while imperfect CSI was considered. We analyzed the secrecy outage performance with different TAS schemes and the closed-form expressions for the lower bound and asymptotic EST with different TAS were obtained. The effect of TAS schemes and parameters of mixed RF-FSO systems on the EST were investigated. The results demonstrated that the proposed TAS schemes can enhance the secrecy performance.
Meanwhile, comparing our results with the results in \cite{Lei2017PJ,Lei2018TCOM, Lei2018PJ}, one can observe that the imperfect CSI seriously deteriorates the secrecy performance of mixed RF-FSO systems. Hereafter, more attention must be given to enhance the secrecy performance of mixed RF-FSO systems with imprecise CSI. One interesting future topic is to investigate to enhance the mixed system with some new technology, such as index modulation technology \cite{MaoT2018Globe}, \cite{LeeY2017Access}.
{In our future work, the scenario that eavesdropper is located in the divergence region of FSO link will be considered and the secrecy performance such that the eavesdropper receives both RF and FSO signals from the $S$ and $R$ will be investigated.}

\section*{Appendix A}
\label{sec:AppendixA}
The PDF of ${h_{RD}}$ is given as \cite{Ansari2016TWC}
\begin{equation}
{f_{{h_{RD}}}}\left( x \right) = \frac{{{\xi ^2}{A_D}}}{{2x}}\sum\limits_{h = 1}^\beta  {{b_h}G_{1,3}^{3,0}\left[ {{\delta _0}x\left| {_{{\xi ^2},\alpha ,h}^{{\xi ^2} + 1}} \right.} \right]} ,
\label{fhRD}
\end{equation}
where
${A_D} = \frac{{2{\alpha ^{0.5\alpha }}}}{{{g^{1 + 0.5\alpha }}\Gamma \left( \alpha  \right)}}{\left( {\frac{{g\beta }}{{g\beta  + {\Omega _1}}}} \right)^{\beta  + 0.5\alpha }}$,
${b_h} = \frac{{\left( {\beta  - 1} \right){{\left( {g\beta  + {\Omega _1}} \right)}^{1 + 0.5\alpha }}{\Omega _1}^{m - 1}}}{{{{\left( {\left( {m - 1} \right)!} \right)}^2}{g^{m - 1}}{\alpha ^{0.5\alpha }}{\beta ^{0.5\alpha  + m}}}}$, and
${\delta _0} = \frac{{\alpha \beta }}{{\left( {g\beta  + \Omega _1} \right){I_l}{A_0}}}$.

Let $X = {\rho _{\text{FSO}}}{h_{RD}} $, $Y = \sqrt {1 - \rho _{\text{FSO}}^2} \varepsilon $, and $Z = {\tilde h_{RD}} $, one can easily obtain
\begin{equation}
\begin{aligned}
{f_X}\left( z \right) = \frac{{{\xi ^2}{A_D}}}{{2z}}\sum\limits_{h = 1}^\beta  {{b_h}G_{1,3}^{3,0}\left[ {\delta z\left| {_{{\xi ^2},\alpha ,h}^{{\xi ^2} + 1}} \right.} \right]},
\label{fX}
\end{aligned}
\end{equation}
\begin{equation}
{f_Y}\left( y \right) = \frac{1}{{\sqrt {2\pi \left( {1 - \rho _{\text{FSO}}^2} \right)} }}{e^{ - \frac{{{y^2}}}{{2\left( {1 - \rho _{\text{FSO}}^2} \right)}}}},
\label{fY}
\end{equation}
where $\delta  = \frac{{\alpha \beta }}{{\left( {g\beta  + \Omega _1} \right){I_l}{A_0}{\rho _{\text{FSO}}}}}$.

According to the convolution theorem, the PDF of $Z$ can be obtained as
\begin{equation}
\begin{aligned}
{f_Z}\left( z \right) &= \int_0^\infty  {{f_X}\left( x \right){f_Y}\left( {z - x} \right)dx} \\
& = \frac{{{\xi ^2}{A_D}}}{{2\sqrt {2\pi \left( {1 - \rho _{\text{FSO}}^2} \right)} }}\sum\limits_{h = 1}^\beta  {{b_h}{e^{ - \frac{{{z^2}}}{{2\left( {1 - \rho _{\text{FSO}}^2} \right)}}}}\nabla },
\label{fZz1}
\end{aligned}
\end{equation}
where
\begin{equation*}
{\nabla  = \int_0^\infty  {\frac{1}{x}{e^{\frac{{zx}}{{1 - \rho _{\text{FSO}}^2}}}}{e^{ - \frac{{{x^2}}}{{2\left( {1 - \rho _{\text{FSO}}^2} \right)}}}}G_{1,3}^{3,0}\left[ {\delta x\left| {_{{\xi ^2},\alpha ,h}^{{\xi ^2} + 1}} \right.} \right]dx} }.
\end{equation*}

Using (11) and (21) of \cite{VS1990}, (9.31.5) of \cite{Gradshteyn2007}, and ${e^x} = \sum\limits_{k = 0}^\infty  {\frac{{{x^n}}}{{n!}}} $, we obtain
\begin{equation}
\nabla  = \frac{{{2^{\alpha -3}}}}{{\pi }}\sum\limits_{k = 0}^\infty  {\frac{{{2^{0.5k + h}}{G_k}}}{{k!{{\left( {1 - \rho _{\text{FSO}}^2} \right)}^{\frac{k}{2}}}}}{z^k}} ,
\label{Delta}
\end{equation}
where ${G_k} = G_{6,3}^{1,6}\left[ {\frac{{{2^3}}}{{{\delta ^2}\left( {1 - \rho _{\text{FSO}}^2} \right)}}\left| {_{\frac{k}{2},\frac{{ - {\xi ^2}}}{2},\frac{{1 - {\xi ^2}}}{2}}^{\frac{{1 - {\xi ^2}}}{2},\frac{{2 - {\xi ^2}}}{2},\frac{{1 - \alpha }}{2},\frac{{2 - \alpha }}{2},\frac{{1 - h}}{2},\frac{{2 - h}}{2}}} \right.} \right]$.

Substituting (\ref{Delta}) into (\ref{fZz1}), we obtain
\begin{equation}
{f_Z}\left( z \right) = B _D\sum\limits_{h = 1}^\beta  {\sum\limits_{k = 0}^\infty  {{H_0}} } e ^{ - \frac{{{z^2}}}{{2\left( {1 - \rho _{\text{FSO}}^2} \right)}}}{z^k},
\label{fZz2}
\end{equation}
where $B _D = \frac{{{\xi ^2}{A_D}{2^{\alpha  - 4.5}}}}{{{\pi ^{1.5}}}}$ and ${H_0} = \frac{{{2^{0.5k + h}}{b_h}{G_k}}}{{k!{{\left( {1 - \rho _{\text{FSO}}^2} \right)}^{\frac{{k + 1}}{2}}}}}$.

It must be noted that the channel gain of FSO link is positive in practical models \cite{FengJ2017OC}. Thus, we rewrite ${f_{{{\tilde h}_D}}}\left( x \right)$ as
\begin{equation}
\begin{aligned}
{f_{{{\tilde h}_{RD}}}}\left( x \right) = \left\{ {\begin{array}{*{20}{c}}
{ {B _D\sum\limits_{h = 1}^\beta  {\sum\limits_{k = 0}^\infty  {{H_0}} } {e ^{ { - \frac{{{x^2}}}{{2\left( {1 - \rho _{\text{FSO}}^2} \right)}}} }}{x^k}} ,x > 0}\\
{1 - {{\text{Z}}_0},\quad \quad \quad \quad \quad \quad \quad \quad \quad x = 0}
\end{array}} \right.,
\label{fhhatD}
\end{aligned}
\end{equation}
where ${{\text{Z}}_0}$ is obtained by $\int_{ 0 }^\infty  {{f_{{{\hat h}_{RD}}}}\left( x \right)dx}  = 1$.
Utilizing \cite[(3.326.2)]{Gradshteyn2007}, we obtain
\begin{equation}
\begin{aligned}
{\text{Z}_0} & = {B_D}\sum\limits_{h = 1}^\beta  {\sum\limits_{k = 0}^\infty  {{H_0}} } \int_0^\infty  {{x^k}{e^{ - \frac{{{x^2}}}{{2\left( {1 - \rho _{{\rm{FSO}}}^2} \right)}}}}dx} \\
& = \frac{{{\xi ^2}{A_D}}}{{{\pi ^{1.5}}}}\sum\limits_{h = 1}^\beta  {\sum\limits_{k = 0}^\infty  {\frac{{{2^{k + \alpha  + h - 5}}{b_h}{G_k}}}{{k!}}\Gamma \left( {\frac{{k + 1}}{2}} \right)} }.
\label{Z0}
\end{aligned}
\end{equation}

Utilizing \cite[(07.34.21.0084.01)]{Wolfram}, we obtain the CDF of ${\tilde h_{RD}} $ as
\begin{equation}
\begin{aligned}
{F_{{{\tilde h}_{RD}}}}\left( x \right) = \left\{ {\begin{array}{*{20}{c}}
  {B _D\sum\limits_{h = 1}^\beta  {\sum\limits_{k = 0}^\infty  {{H_1}} } G_{1,2}^{1,1}\left[ {\frac{{{x^2}}}{{2\left( {1 - \rho _{\text{FSO}}^2} \right)}}\left| {_{\frac{{k + 1}}{2},0}^1} \right.} \right],x > 0} \\
  {1 - {{\text{Z}}_0}, \quad \quad \quad \quad \quad \quad \quad \quad \quad \quad \quad \quad \,\,\, x = 0}
\end{array}} \right.,
\label{FhhatD}
\end{aligned}
\end{equation}
where ${H_1} = \frac{{{b_h}{2^{k + h - 0.5}}{G_k}}}{{k!}}$.

{Although the expression of the CDF in (\ref{FhhatD}) includes the infinite summation, the method utilized here is the same as that in \cite{FengJ2017OC} and the truncation errors versus the number of summation items is given in Fig. 2 of \cite{Lei2018PJ}, which proves that the infinite summation is convergent with a finite truncation. }

Based on (\ref{gammaRD}), we obtain the result in (\ref{pdfgammaRD}) and (\ref{cdfgammaRD}).

\section*{Appendix B}  
\label{sec:AppendixB}
The PDF and CDF of
${\gamma _{{S_i}R}} = \frac{{{P_S}}}{{{\sigma ^2}}}\sum\limits_{j = 1}^{{N_R}}{{{\left|{{h_{{S_i}{R_j}}}}\right|}^2}} $
are given by \cite{Lei2017TVT}
\begin{equation}
{f_{{\gamma _{{S_i}R}}}}\left( x \right) = \frac{{{\lambda _R^{{\tau _R}}}}}{{\Gamma \left( {{\tau _R}} \right)}}{e^{ - {\lambda _R}x}}{x^{{\tau _R} - 1}},
\label{fgammasir}
\end{equation}
\begin{equation}
{F_{{\gamma _{{S_i}R}}}}\left( x \right) = 1 - \sum\limits_{i = 0}^{{\tau _R} - 1} {\frac{{{e^{ - {\lambda _R}x}}}}{{i!}}{{\left( {{\lambda _R}x} \right)}^i}},
\label{Fgammasir1}
\end{equation}
respectively, where ${\lambda _R} = \frac{{{m_R}}}{{{{\bar \gamma }_R}}}$, ${\tau _R} = {m _R}{N _R}$, and ${\bar \gamma _R}$ is the average SNR of $S - R$ link.

The PDF of ${\hat \gamma _{SR}}$ can be obtained by \cite{Ferdinand2013CL}
\begin{equation}
\begin{aligned}
{f_{{{\hat \gamma }_{SR}}}}\left( x \right) &= \int_0^\infty  {{f_{{{\hat \gamma }_{SR}}\left| {{\gamma _{SR}}} \right.}}\left( {x\left| y \right.} \right){f_{{\gamma _{SR, 1}}}}\left( y \right)dy}  \\
& = \int_0^\infty  {\frac{{{f_{{{\hat \gamma }_{{S_i}R}},{\gamma _{{S_i}R}}}}\left( {x,y} \right)}}{{{f_{{\gamma _{{S_i}R}}}}\left( y \right)}}{f_{{\gamma _{SR, 1}}}}\left( y \right)dy},
\label{fhatgammaSR}
\end{aligned}
\end{equation}
where ${\gamma _{SR, 1}} = \mathop {\max }\limits_{1 \le i \le {N_S}} \left\{ {{\gamma _{{S_i}R}}} \right\}$ and the joint PDF ${f_{{{\hat \gamma }_{{S_i}R}},{\gamma _{{S_i}R}}}}\left( {x,y} \right)$ is given by \cite{Simon2000Book}
\begin{equation}
\begin{aligned}
{f_{{{\hat \gamma }_{{S_i}R}},{\gamma _{{S_i}R}}}}\left( {x,y} \right) &= \frac{{\lambda _R^{{\tau _R} + 1}}}{{\left( {1 - \rho _{SR}^2} \right)\Gamma \left( {{\tau _R}} \right)}}{\left( {\frac{{xy}}{{{\rho _{SR}^2}}}} \right)^{\frac{{{\tau _R} - 1}}{2}}}\\
& \times {e^{ - \frac{{\left( {x + y} \right){\lambda _R}}}{{1 - \rho _{SR}^2}}}}{I_{{\tau _R} - 1}}\left( {\frac{{2{\lambda _R\rho _{SR}}\sqrt {xy} }}{{1 - \rho _{SR}^2}}} \right),
\label{joint}
\end{aligned}
\end{equation}
where ${I_n}\left( x \right)$ is the $n$-th order modified Bessel function of first kind, defined by \cite[(8.406.1)]{Gradshteyn2007}.

Based on ${\gamma _{SR, 1}} = \mathop {\max }\limits_{1 \le i \le {N_S}} \left\{ {{\gamma _{{S_i}R}}} \right\}$, we obtain the PDF of ${\gamma _{SR, 1}} $ as
\begin{equation}
\begin{aligned}
{f_{{\gamma _{SR, 1}}}}\left( x \right) &= {N_S}{\left( {{F_{{\gamma _{{S_i}R}}}}\left( x \right)} \right)^{{N_S} - 1}}{f_{{\gamma _{{S_i}R}}}}\left( x \right)\\
& = {N_S}{\left( {1 - \sum\limits_{i = 0}^{{\tau _R} - 1} {\frac{{{e^{ - {\lambda _R}x}}}}{{i!}}{{\left( {{\lambda _R}x} \right)}^i}} } \right)^{{N_S} - 1}}{f_{{\gamma _{{S_i}R}}}}\left( x \right).
\label{fgammaSR0}
\end{aligned}
\end{equation}
Making using of multinomial theorem, namely,
\begin{equation}
{\left( {\sum\limits_{i = 1}^m {{x_i}} } \right)^N} = \sum\limits_{{n_1} + {n_2} + ... + {n_m} = N} {\frac{{N!}}{{\prod\limits_{j = 1}^m {\left( {{n_j}} \right)!} }}\prod\limits_{k = 1}^m {{{\left( {{x_k}} \right)}^{{n_k}}}} },
\label{duoxiangshi}
\end{equation}
where ${n_i}\left( {i = 1, \cdots ,m} \right) \in \mathbb{N}$, which denotes a non-negative integer set. We expressed the PDF of ${\gamma _{SR, 1}} $ as
\begin{equation}
{f_{{\gamma _{SR, 1}}}}\left( x \right) = \frac{{{N_S}{\lambda _R^{{\tau _R}}}}}{{\Gamma \left( {{\tau _R}} \right)}}\sum\limits_{S _R} {{A _R}{x^{{B _R } + {\tau _R} - 1}}{e^{ - \lambda _R \left( {{C _R} + 1} \right)x}}},
\label{fgammaSR1}
\end{equation}
where ${S_R} = \left\{ {\left( {{n_1}, \cdots ,{n_{{\tau _R} + 1}}} \right) \in \mathbb{N}\left| {\sum\limits_{p = 1}^{{\tau _R} + 1} {{n_p} = {N_S} - 1} } \right.} \right\}$,
${A_R} = \left( {\frac{{{N_S} - 1}}{{\prod\limits_{q = 1}^{{\tau _R} + 1} {{n_q}} }}} \right)\prod\limits_{p = 2}^{{\tau _R} + 1} {{{\left( { - \frac{{\lambda _R^{p - 2}}}{{\left( {p - 2} \right)!}}} \right)}^{{n_p}}}} $,
${B _R} = \sum\limits_{p = 2}^{{\tau _R} + 1} {{n_p}\left( {p - 2} \right)} $, and
${C _R} = \sum\limits_{p = 2}^{{\tau _R} + 1} {{n_p}} $.

Substituting (\ref{joint}) and (\ref{fgammaSR1}) into (\ref{fhatgammaSR}), we have
\begin{equation}
\begin{aligned}
{f_{{{\hat \gamma }_{SR}}}}\left( x \right) &= \frac{{{N_S}{\lambda _R^{{\tau _R} + 1}}}}{{\left( {1 - \rho _{SR}^2} \right)\Gamma \left( {{\tau _R}} \right)}}{\left( {\frac{x}{{{\rho _{SR}^2}}}} \right)^{\frac{{{\tau _R} - 1}}{2}}}\sum\limits_{S _R} {{A _R}{e^{ - \frac{{{\lambda _R}x}}{{1 - \rho _{SR}^2}}}}} \\
& \times \int_0^\infty  {{y^{{B _R} + \frac{{{\tau _R} - 1}}{2}}}{e^{ - {\chi _1}y}}{I_{{\tau _R} - 1}}\left( {2{\chi _2}\sqrt {xy} } \right)dy},
\end{aligned}
\end{equation}
where ${\chi _1} = \lambda _R\left( {\frac{1}{{1 - \rho _{SR}^2}} + C _ R} \right)$ and ${\chi _2} = \frac{{\lambda _R {{\rho _{SR}}} }}{{1 - \rho _{SR}^2}}$.

Making use of (8.406.3), (6.643.4), and (8.970.1) of \cite{Gradshteyn2007}, and after some algebraic manipulations, we obtain (\ref{pdfgammasrhat}) and (\ref{cdfgammasrhat}), respectively.

\section*{Appendix C}  
\label{sec:AppendixC}
The PDF and CDF of ${\gamma _{{S_i}E}} = \frac{{{P_S}}}{{{\sigma ^2}}}\sum\limits_{j = 1}^{{N_E}} {{{\left| {{h_{{S_i}{E_j}}}} \right|}^2}} $
are given by \cite{Lei2017TVT}
\begin{equation}
{f_{{\gamma _{{S_i}E}}}}\left( x \right) = \frac{{{\lambda _E^{{\tau _E}}}}}{{\Gamma \left( {{\tau _E}} \right)}}{e^{ - {\lambda _E}x}}{x^{{\tau _E} - 1}},
\label{fgammasie2}
\end{equation}
\begin{equation}
{F_{{\gamma _{{S_i}E}}}}\left( x \right) = 1 - \sum\limits_{i = 0}^{{\tau _E} - 1} {\frac{{{e^{ - {\lambda _E}x}}}}{{i!}}{{\left( {{\lambda _E}x} \right)}^i}},
\label{Fgammasie2}
\end{equation}
respectively, where ${\lambda _E} = \frac{{{m_E}}}{{{{\bar \gamma }_E}}}$, ${\tau _E} = {m _E}{N _E}$, and ${\bar \gamma _E}$ is the average SNR of $S$-$E$ link.

The PDF of ${\hat \gamma _{SE}}$ can be obtained by \cite{Ferdinand2013CL}
\begin{equation}
\begin{aligned}
{f_{{{\hat \gamma }_{SE}}}}\left( x \right) &= \int_0^\infty  {{f_{{{\hat \gamma }_{SE}}\left| {{\gamma _{SE}}} \right.}}\left( {x\left| y \right.} \right){f_{{\gamma _{SE, 2}}}}\left( y \right)dy}  \\
& = \int_0^\infty  {\frac{{{f_{{{\hat \gamma }_{{S_i}E}},{\gamma _{{S_i}E}}}}\left( {x,y} \right)}}{{{f_{{\gamma _{{S_i}E}}}}\left( y \right)}}{f_{{\gamma _{SE, 2}}}}\left( y \right)dy},
\label{fhatgammaSR2}
\end{aligned}
\end{equation}
where ${\gamma _{SE, 2}} = \mathop {\min }\limits_{1 \le i \le {N_S}} \left\{ {{\gamma _{{S_i}E}}} \right\}$ and the joint PDF ${f_{{{\hat \gamma }_{{S_i}E}},{\gamma _{{S_i}E}}}}\left( {x,y} \right)$ is given by \cite{YangN2012TCOM}
\begin{equation}
\begin{aligned}
{f_{{{\hat \gamma }_{{S_i}E}},{\gamma _{{S_i}E}}}}\left( {x,y} \right) &= \frac{{\lambda _E^{{\tau _E} + 1}}}{{\left( {1 - \rho _{SE}^2} \right)\Gamma \left( {{\tau _E}} \right)}}{\left( {\frac{{xy}}{{{\rho _{SE}^2}}}} \right)^{\frac{{{\tau _E} - 1}}{2}}}\\
& \times {e^{ - \frac{{\left( {x + y} \right){\lambda _E}}}{{1 - \rho _{SE}^2}}}}{I_{{\tau _E} - 1}}\left( {\frac{{2{\lambda _E{\rho _{SE}}}\sqrt {xy} }}{{1 - \rho _{SE}^2}}} \right).
\label{jointse}
\end{aligned}
\end{equation}

Then we obtain the PDF of ${\gamma _{SE, 2}} $ as
%
\begin{equation}
\begin{aligned}
{f_{{\gamma _{SE, 2}}}}\left( x \right) &= {N_S}{\left( {1 - {F_{{\gamma _{{S_i}E}}}}\left( x \right)} \right)^{{N_S} - 1}}{f_{{\gamma _{{S_i}E}}}}\left( x \right)\\
& = \frac{{{N_S}{\lambda _E^{{\tau _E}}}}}{{\Gamma \left( {{\tau _E}} \right)}}\sum\limits_{{S_E}} {{A_E}{x^{{B_E} + {\tau _E} - 1}}{e^{ - {\lambda _E}\left( {{C_E} + 1} \right)x}}} ,
\label{fgammaSR}
\end{aligned}
\end{equation}
where ${S _E} = \left\{ {\left( {{n_1}, \cdots, {n_{{\tau _E} }}} \right) \in \mathbb{N}\left| {\sum\limits_{p = 1}^{{\tau _E}} {{n_p} = {N_S} - 1} } \right.} \right\}$,
${A_E} = \left( {\frac{{{N_S} - 1}}{{\prod\limits_{q = 1}^{{\tau _E}} {{n_q}} }}} \right)\prod\limits_{p = 1}^{{\tau _E}} {{{\left( {\frac{{\lambda _E^{p - 1}}}{{\left( {p - 1} \right)!}}} \right)}^{{n_p}}}} $,
${B_E} = \sum\limits_{p = 1}^{{\tau _E}} {{n_p}\left( {p - 1} \right)} $, and
${C_E} = \sum\limits_{p = 1}^{{\tau _E}} {{n_p}} $.

Substituting (\ref{joint}) and (\ref{fgammaSR}) into (\ref{fhatgammaSR}), we have
\begin{equation}
\begin{aligned}
{f_{{{\hat \gamma }_{SE}}}}\left( x \right) &= \frac{{{N_S}{\lambda _E^{{\tau _E} + 1}}}}{{\left( {1 - \rho _{SE}^2} \right)\Gamma \left( {{\tau _E}} \right)}}{\left( {\frac{x}{{{\rho _{SE}^2}}}} \right)^{\frac{{{\tau _E} - 1}}{2}}}\sum\limits_{S _E} {{A _E}{e^{ - \frac{{{\lambda _E}x}}{{1 - \rho _{SE}^2}}}}} \\
& \times \int_0^\infty  {{y^{{B _E} + \frac{{{\tau _E} - 1}}{2}}}{e^{ - {\chi _3}y}}{I_{{\tau _E} - 1}}\left( {2{\chi _4}\sqrt {xy} } \right)dy},
\end{aligned}
\end{equation}
where ${\chi _3} = \lambda _R\left( {\frac{1}{{1 - \rho _{SE}^2}} + C _ E} \right)$ and ${\chi _4} = \frac{{\lambda _E {{\rho _{SE}}} }}{{1 - \rho _{SE}^2}}$.

Making use of (8.406.3), (6.643.4), and (8.970.1) of \cite{Gradshteyn2007}, and after some algebraic manipulations, we obtain (\ref{pdfgammasehat}) and (\ref{cdfgammasehat}), respectively.

\section*{Acknowledgment}
The authors would like to thank the Editor for efficiently handling the review of this paper and the anonymous reviewers for their valuable suggestions and comments that helped
to improve the quality of the paper.

\end{document}